\documentclass[sigconf]{acmart}
\AtBeginDocument{%
  \providecommand\BibTeX{{%
    \normalfont B\kern-0.5em{\scshape i\kern-0.25em b}\kern-0.8em\TeX}}}

\usepackage{color,soul}
\usepackage{booktabs} %much nicer tables
\usepackage{subfig} %for putting multiple subimages in one figure, with individual captions
\usepackage{graphicx}
\usepackage[utf8]{inputenc}
\usepackage[T1]{fontenc}

\usepackage{siunitx}
\newcommand{\hide}[1]{}

\sethlcolor{yellow}
\ifdefined\scifihidecomments
    \newcommand{\cheng}[1] {}
    \newcommand{\hyunc}[1] {} % Hyunchul
    \newcommand{\yax}[1] {} % Yaxuan
    \newcommand{\tuoc}[1] {} % Tuochao
    \newcommand{\sony}[1] {} % Songyun
    \newcommand{\ReviewerFeedback}[1] {}  
    \newcommand{\fran}[1] {}
    \newcommand{\rz}[1]{}
    \newcommand{\lw}[1] {}
    \newcommand{\mose}[1] {} % Mose
    \newcommand{\ke}[1] {} % Ke
\else
    \definecolor{burntorange}{rgb}{0.8, 0.33, 0.0}
    \definecolor{cadmiumgreen}{rgb}{0.0, 0.42, 0.24}
    \definecolor{cobalt}{rgb}{0.0, 0.28, 0.67}
    \definecolor{amber}{rgb}{1.0, 0.75, 0.0}
    \definecolor{fashionfuchsia}{rgb}{0.96, 0.0, 0.63}
    \definecolor{brightcerulean}{rgb}{0.11, 0.67, 0.84}
    \definecolor{frenchblue}{rgb}{0.0, 0.45, 0.73}
    \definecolor{darkslateblue}{rgb}{0.28, 0.24, 0.55}
    \definecolor{cerulean}{rgb}{0.0, 0.48, 0.65}
    \definecolor{darkpastelgreen}{rgb}{0.01, 0.75, 0.24}
    \newcommand{\hyunc}[1] { \textcolor{burntorange}{[{\hl{hyunc:}} {#1}}]}
    \newcommand{\yax}[1] { \textcolor{magenta}{[{\hl{yaxuan:}} {#1}]}}
    \newcommand{\tuoc}[1] { \textcolor{darkpastelgreen}{[{\hl{tuochao:}} {#1}]}}
    \newcommand{\sony}[1] { \textcolor{blue}{[{\hl{songyun:}} {#1}]}}
    \newcommand{\ReviewerFeedback}[1] { \textcolor{brightcerulean}{[{Reviewer Feedback:} {#1}}]}
    \newcommand{\fran}[1]{\textcolor{burntorange}{[{francois:}{#1}}]}
    \newcommand{\rz}[1]{\textcolor{teal}{[{Ruidong: }{#1}]}}
    \newcommand{\lw}[1]{\textcolor{fashionfuchsia}{[{liuwei:}{#1}}]}
    \newcommand{\mose}[1]{\textcolor{burntorange}{[{mose:}{#1}}]}
    \newcommand{\ke}[1] { \textcolor{red!55!yellow}{[{Ke:} {#1}}]}

\fi

\ifdefined\scifiblind
    \newcommand{\blind}[1]{[omitted for blind review]}
\else
    \newcommand{\blind}[1]{#1} %for camera ready (not blinded)
\fi

\usepackage{siunitx}
\usepackage{subfloat}
\usepackage{multirow}
\usepackage{pifont}
\usepackage{gensymb}
\newcommand{\cmark}{\text{\ding{51}}}
\newcommand{\xmark}{\text{\ding{55}}}
\copyrightyear{2024}
\acmYear{2024}
\setcopyright{acmlicensed}\acmConference[CHI '24]{Proceedings of the CHI Conference on Human Factors in Computing Systems}{May 11--16, 2024}{Honolulu, HI, USA}
\acmBooktitle{Proceedings of the CHI Conference on Human Factors in Computing Systems (CHI '24), May 11--16, 2024, Honolulu, HI, USA}
\acmDOI{10.1145/3613904.3642613}
\acmISBN{979-8-4007-0330-0/24/05}

\begin{document}

% \title{EarIO: a low-power, minimally-obtrusive and practical acoustic sensing method to track continuous full facial expressions using subtle skin deformations}

\title{EyeEcho: Continuous and Low-power Facial Expression Tracking on Glasses}
\author{Ke Li}
\affiliation{%
  \institution{Cornell University}
  \city{Ithaca}
  \country{USA}}
\email{kl975@cornell.edu}
\orcid{0000-0002-4208-7904}

\author{Ruidong Zhang}
\affiliation{%
  \institution{Cornell University}
  \city{Ithaca}
  \country{USA}}
\email{rz379@cornell.edu}
\orcid{0000-0001-8329-0522}

\author{Siyuan Chen}
\affiliation{%
  \institution{Cornell University}
  \city{Ithaca}
  \country{USA}}
\email{sc2489@cornell.edu}
\orcid{0009-0008-4828-3392}

\author{Boao Chen}
\affiliation{%
  \institution{Cornell University}
  \city{Ithaca}
  \country{USA}}
\email{bc526@cornell.edu}
\orcid{0000-0002-3527-9481}

\author{Mose Sakashita}
\affiliation{%
  \institution{Cornell University}
  \city{Ithaca}
  \country{USA}}
\email{ms3522@cornell.edu}
\orcid{0000-0003-4953-2027}

\author{François Guimbretière}
\affiliation{%
  \institution{Cornell University}
  \city{Ithaca}
  \country{USA}}
\email{fvg3@cornell.edu}
\orcid{0000-0002-5510-6799}

\author{Cheng Zhang}
\affiliation{%
  \institution{Cornell University}
  \city{Ithaca}
  \country{USA}}
\email{chengzhang@cornell.edu}
\orcid{0000-0002-5079-5927}

\renewcommand{\shortauthors}{Li et al.}

\begin{abstract}
In this paper, we introduce EyeEcho, a minimally-obtrusive acoustic sensing system designed to enable glasses to continuously monitor facial expressions. It utilizes two pairs of speakers and microphones mounted on glasses, to emit encoded inaudible acoustic signals directed towards the face, capturing subtle skin deformations associated with facial expressions. The reflected signals are processed through a customized machine-learning pipeline to estimate full facial movements. EyeEcho samples at 83.3 Hz with a relatively low power consumption of $167 mW$. Our user study involving 12 participants demonstrates that, with just four minutes of training data, EyeEcho achieves highly accurate tracking performance across different real-world scenarios, including sitting, walking, and after remounting the devices. Additionally, a semi-in-the-wild study involving 10 participants further validates EyeEcho's performance in naturalistic scenarios while participants engage in various daily activities. Finally, we showcase EyeEcho's potential to be deployed on a commercial-off-the-shelf (COTS) smartphone, offering real-time facial expression tracking.

\end{abstract}







\maketitle

\section{Introduction}

Facial expressions play an important role in various interaction applications including video calls, facial gesture input~\cite{rantanen2010capacitive,lankes2008facial,matthies2017earfieldsensing}, non-verbal communications (e.g. sign language~\cite{Agris2008significance}) and are indispensable in virtual environments. Continuous and accurate facial expression tracking is critical for an immersive and convenient interaction experience for users. Deploying such technologies on eyewears such as smart glasses (Lenovo ThinkReality A3~\cite{ThinkReality} and Bose Frames Tempo~\cite{bosetempo}) and augmented reality (AR) glasses (Google Glass~\cite{googleglass} and Xreal Air 2 Ultra~\cite{xreal}) is especially promising and important as these devices are widely available and serve as direct media in these interaction applications.

\begin{figure*}[t]
  \includegraphics[width=1 \textwidth]{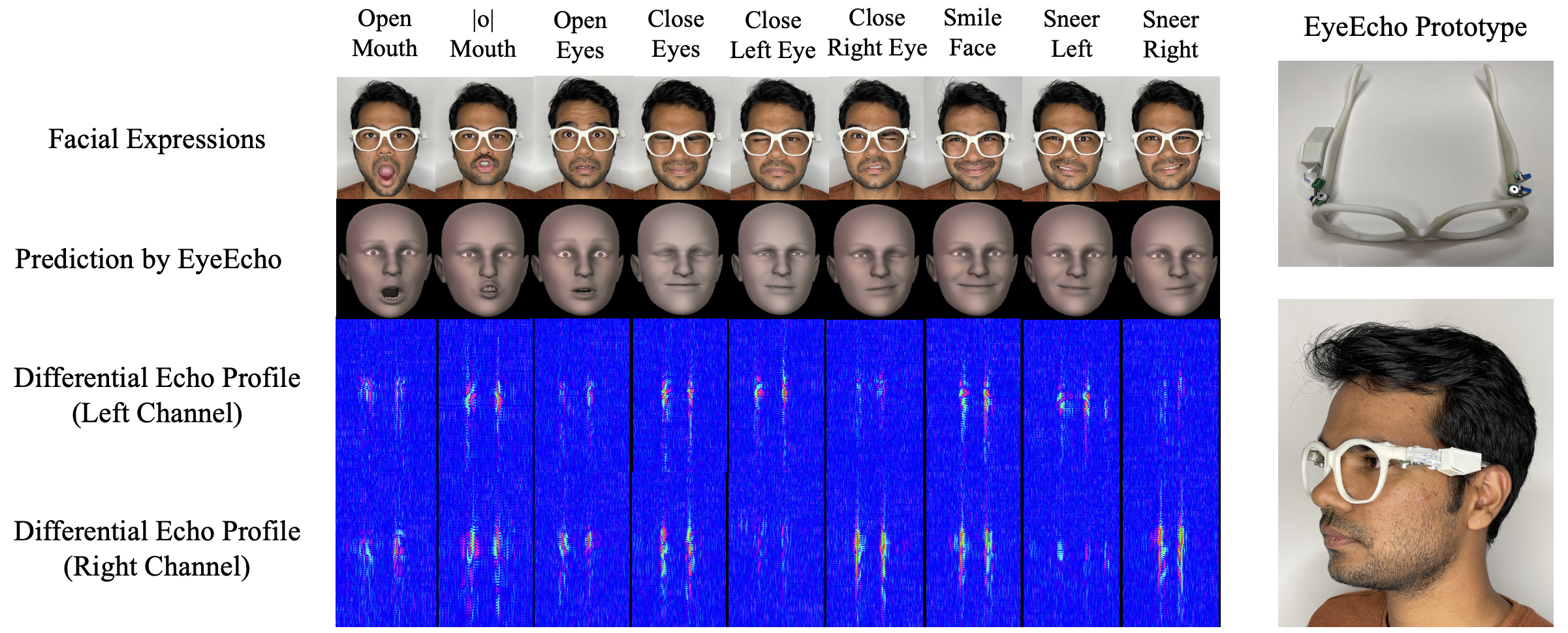}
  \caption{Designed Facial Expressions and Corresponding Differential Echo Profiles.}
  \label{Fig: Facial Expressions with Different Echo Profiles}
  \Description{This figure shows the nine facial expressions used in our study performed by one person wearing the glasses form factor embedded with our EyeEcho system, which are Open Mouth, |o| Mouth, Open Eyes, Close Eyes, Close Left Eye, Close Right Eye, Smile Face, Sneer Left, and Sneer Right. Under the nine facial expressions, the figure shows an avatar with these nine facial expressions which are predicted and generated by our EyeEcho system. Then under these are nine figures of different echo profiles corresponding to these nine facial expressions for both left and right channels of microphones. The differential echo profiles display different patterns for different facial expressions and are used for model prediction. On the right of the figure, a picture of our EyeEcho device and a person wearing it are shown.}
\end{figure*}

However, developing continuous facial expression tracking technologies on glasses presents unique challenges due to the constraints of the sensor size and battery capacity. First, capturing both upper and lower facial movements is crucial, but glasses primarily cover the upper face, making lower face tracking challenging. Maintaining high temporal and spatial resolutions of continuous facial expression tracking is also difficult. Existing glasses-based methods can only recognize discrete facial gestures. For instance, recent work on glasses using the acoustic-based method can only recognize 6 discrete upper facial expressions \cite{xie2021acoustic}. Second, smart glasses often come with limited battery capacity due to the weight restriction. Power-hungry sensors like RGB cameras drain batteries on smart glasses quickly, e.g., in less than one hour~\cite{paro2015video}. Third, the need for reliable performance after glasses are remounted (taken off and taken back on) poses challenges, especially for on-skin sensors (e.g. EMG sensors). As a result, continuous facial expression tracking on glasses has not been explored extensively. Other wearables like necklaces \cite{chen2021n} and earpieces \cite{li2022eario,chen2020c,wu2021bioface} have been used, but they come with their own limitations. Earphones, for instance, face issues with sensor placement and when tracking subtle upper-face movements. Despite many people wearing glasses regularly, \textit{there is no existing technology on glasses that can continuously track both upper and lower facial expressions}.

The research question of this paper is: \textit{Can we develop a sensing system on glasses to track facial expressions continuously that are light-weight, low-power, and can work well in real-world settings? }
To answer this research question, we present EyeEcho, an intelligent acoustic sensing solution that can continuously track high-resolution facial expressions on glasses by analyzing the skin deformations around the cheeks captured using only two pairs of on-device speaker and microphone with inaudible acoustic signals (Frequency Modulated Continuous Wave, FMCW). A customized convolutional neural network is developed to estimate facial expressions represented by 52 blend-shape parameters calculated using Apple's ARKit API~\cite{arkit} from the processed acoustic signals (echo profiles). A user study with 12 participants showed that EyeEcho can accurately estimate facial expressions continuously on glasses using only four minutes of training data from each participant. Besides, it is able to detect eye blinks with an F1 score of 82\%, which has not been shown in any of the prior work \cite{li2022eario,wu2021bioface,chen2020c,chen2021n}. 

In order to gain a deeper understanding of the EyeEcho's performance in real-world scenarios, we conducted a semi-in-the-wild study involving 10 participants to evaluate the performance of the facial expression tracking system on glasses in a naturalistic setting. This study took place in three distinct rooms of an apartment: the living room, bedroom, and kitchen. The study aimed to assess the system's ability to track facial expressions continuously while participants engaged in various common daily activities. Importantly, the study encompassed real-world ambient noises, such as the sound of videos, noise from the microwave oven, and the hum of the refrigerator among others. In total, around 700 minutes of data were collected in this study. The results demonstrate that our system still showed reliable performance in tracking facial expressions throughout the participants' engagement in diverse activities across different rooms and days.

In addition to the promising tracking performance, EyeEcho is also relatively low-power and light-weight compared to camera-based facial expression tracking technologies on wearables. The full sensing system including sensors, Bluetooth module and microprocessors can operate at a sample rate of $83.3 Hz$ with a power signature of $167 mW$. Theoretically, it can last for around 14 hours using the battery of Google Glass ($570 mAh$)~\cite{googleglass}, if EyeEcho is used alone. With the usage of more power-efficient speakers, the power consumption of our system can be optimized to as low as $71 mW$. The ML algorithm is optimized to be lightweight based on the ResNet-18 architecture, so that it can be deployed on a commodity smartphone for real-time processing. EyeEcho is able to track users' facial expressions continuously at $29 Hz$ in real-time on a commodity Android phone, which was not shown in any of the similar sensing systems \cite{li2022eario,chen2020c,chen2021n,wu2021bioface}. We believe EyeEcho has significantly advanced the field of tracking facial expressions on glasses by offering a low-power and minimally-obtrusive sensing solution that can be deployed on commodity smartphones for real-time tracking. The key contributions of this paper are summarized below:

\begin{itemize}
    \item Enabled continuous facial expression tracking on glasses using low-power acoustic sensing;   
    \item Conducted studies, including a semi-in-the-wild study, to evaluate EyeEcho in estimating facial expressions including eye blinks in both lab and real-world settings; 
    \item Developed a real-time processing system on an Android phone and demonstrated promising performance. 
    
\end{itemize}

\section{Related Work}
\label{Sec: Related Work}

\begin{table*}[t]
\caption{EyeEcho and Other Continuous Facial Expression Tracking Wearables. Power consumption only includes the data collection and transmission (if any) unit. NS = Not Specified.}
\label{Tab: comparision with previous}
\small
\begin{tabular}{| c | c | c | c | c | c | c | c | c | c |} 
\hline
 \multirow{3}*{Project} & \multirow{3}*{Form Factor} & \multirow{3}*{Sensors} & \multirow{3}*{Power} & Training & Evaluated & Evaluated & Evaluated & Blinking & Deployed on\\
 ~ & ~ & ~ & ~ & Data & across & while & Semi-in- & Detection & Phone in\\
 ~ & ~ & ~ & ~ & Needed & Sessions? & Walking? & the-Wild? & Included? & Real-time?\\ [0.5ex]
 \hline\hline
 \textbf{EyeEcho} & \textbf{Glasses} & \textbf{Acoustics} & \textbf{167mW} & \textbf{4 mins} & \textbf{\cmark} & \textbf{\cmark} & \textbf{\cmark} & \textbf{\cmark} & \textbf{\cmark}\\
 \hline
 EarIO~\cite{li2022eario} & Earphones & Acoustics & 154mW & 32 mins & \cmark & \cmark & \xmark & \xmark & \xmark\\
 \hline
 EARFace~\cite{zhang2023i} & Earphones & Acoustics & 245mW & 2 mins & \cmark & \cmark & \xmark & \cmark & \cmark\\
 \hline
 NeckFace~\cite{chen2021n} & Neck-lace/-band & Cameras & 4W & 7 mins & \cmark & \cmark & \xmark & \xmark & \xmark\\
 \hline
 C-Face~\cite{chen2020c} & Ear-/Head-phone & Cameras & >4W & 6 mins & \cmark & \xmark & \xmark & \xmark & \xmark\\
 \hline
 BioFace-3D~\cite{wu2021bioface} & Single Earpiece & Biosensors & 138mW & 20 mins & \xmark & \xmark & \xmark & \xmark & \xmark\\
 \hline
%  Xie et al.~\cite{xie2021acoustic} & Glasses & Acoustics & NS & 6 Expressions & F1-score: 0.92\\
%  \hline
 Wei et al.~\cite{wei2019vr} & VR Headset & Cameras & NS & NS & \cmark & \xmark & \xmark & \cmark & \xmark\\
 \hline
 Pantœnna~\cite{kim2023pan} & VR Headset & Antenna & \textasciitilde 1W & \textasciitilde 16 mins & \cmark & \xmark & \xmark & \xmark & \xmark\\
 \hline
\end{tabular}
\end{table*}

\subsection{Non-wearable Facial Expression Tracking}
\label{Subsec: related work no-wearable}
The most commonly used non-wearable technologies to track facial expressions are based on cameras placed in front of the user to capture the face. 
Researchers rely on the images captured by RGB cameras~\cite{hsieh2015unconstrained,thies2015real}, thermal infrared cameras~\cite{he2013facial}, and/or depth cameras~\cite{hsieh2015unconstrained,ijjina2014facial,thies2015real}, or images from existing datasets~\cite{kahou2013combining,liu2014learning,liu2014facial,rifai2012disentangling,wu2018look,sebe2007authentic} to develop algorithms to track subjects' facial expressions. Recently, learning-based algorithms show impressive performance on tracking facial expressions with the support of fast-developing deep learning models, such as Convolutional Neural Network (CNN)~\cite{ijjina2014facial,kahou2013combining,rifai2012disentangling,wu2018look}, Generative Adversarial Network (GAN)~\cite{lai2018emotion}, Deep Belief Network (DBN)~\cite{he2013facial,kahou2013combining,liu2014facial,ranzato2011deep}, etc. Due to their impressive performance of tracking facial expressions and minimum requirement of the amount of training data needed, frontal-camera-based algorithms have been used as the ground truth acquisition methods in many wearable facial expression tracking systems, with the help of several public libraries, e.g., the Dlib library~\cite{king2009dlib} and the Apple's ARKit API~\cite{arkit}. 
Despite their satisfactory tracking performance, these technologies usually require capturing the full face in the image, which does not work well when the user's face is occluded. Some researchers have put efforts in reconstructing users' facial expressions when part of their face is occluded~\cite{hsieh2015unconstrained,lai2018emotion,wu2018look}. However, the frontal-camera-based algorithms are still easily impacted by lighting conditions in the environment and does not work well while users are in motion.

Apart from frontal camera based methods, recently some researchers placed speakers and microphones in front of the user to recognize facial expressions and emotional gestures~\cite{gao2022sonicface}. This alleviated some privacy concerns brought by camera-based methods, but the system can only recognize discrete expressions and emotions. In the meantime, other researchers also use frontal speakers and microphones to detect more subtle facial movements of the user, which are blinking~\cite{liu2021blinklistener}. mm3DFace~\cite{xie2023mm3dface} reconstructed users' facial expressions continuously with a competitive performance by placing a millimeter
wave (mmWave) radar in front of them. These non-camera-based technologies demonstrate the potential of sensing modalities other than cameras for tracking facial movements but still suffer from the limitations of non-wearable devices that they do not work well when the user's face is occluded or the user is walking around.

\subsection{Wearable Facial Expression Tracking}
\label{Subsec: related work wearables}

\subsubsection{Smart Glasses}

Several prior projects implemented facial expression recognition on glasses~\cite{scheirer1999expression,masai2016facial,Hu2020Demo,kwon2021emotion,xie2021acoustic}, using a variety of sensors, including piezoelectric sensors~\cite{scheirer1999expression}, photo reflective sensors~\cite{masai2016facial}, cameras~\cite{Hu2020Demo,kwon2021emotion}, biosensors~\cite{kwon2021emotion}, speakers and microphones~\cite{xie2021acoustic}. However, all of them are only capable of distinguishing several discrete facial expressions. To the best of our knowledge, we have not seen any prior work that can track full facial movements continuously on glasses. Usually, glasses are small and lightweight, thus having a limited battery and processor. Therefore, it places a high demand for the sensing technology on its size, weight and energy consumption.

\subsubsection{Other Wearables}
Other wearables designed to track facial expressions include ear-mounted devices using cameras~\cite{chen2020c}, speakers and microphones~\cite{li2022eario,zhang2023i}, EMG or/and electrooculography~(EOG) sensors~\cite{gruebler2010measurement,wu2021bioface}, 
Inertial Measurement Unit~(IMU) sensors~\cite{verma2021e}, or barometers~\cite{ando2017canalsense}, face mask using acoustic signals~\cite{iravantchi2019interferi}, necklace/neckband using cameras~\cite{chen2021n}, and VR headsets using cameras~\cite{wei2019vr} or an antenna~\cite{kim2023pan}. However, most of the work are only able to recognize discrete facial gestures. Six recent work, C-Face~\cite{chen2020c}, NeckFace~\cite{chen2021n}, BioFace-3D~\cite{wu2021bioface}, EarIO~\cite{li2022eario}, Pantœnna~\cite{kim2023pan}, and EARFace~\cite{zhang2023i}, show the ability to track facial expressions continuously on wearables, which we will compare in detail in the next subsubsection.

\subsubsection{Comparison between EyeEcho and Prior Work}
\label{Subsec: comparison with prior work}
We summarize continuous facial expression tracking technologies on wearables that are closest to our work in Tab.~\ref{Tab: comparision with previous}. Compared with previous work, EyeEcho excels in at least one of the following aspects: 1) tracking capability (continuous vs. discrete, 3D blendshapes vs. 2D landmarks), 2) obtrusiveness, 3) power consumption and 4) performance. For example, C-Face~\cite{chen2020c} and NeckFace~\cite{chen2021n} use cameras as the sensing unit, which leads to high energy consumption. NeckFace operates at $4 W$ which is 24 times higher than our system. Pantœnna~\cite{kim2023pan} instrumenting the VR headset with an antenna to emit RF signals also consumes power as high as $1 W$. Furthermore, Pantœnna can only track mouth, i.e. lower-face, movements and the size of the antenna makes it difficult to be deployed on glasses which are smaller and more lightweight. BioFace-3D~\cite{wu2021bioface} does a great job in maintaining lower-power but its device requires attaching electrodes of biosensors onto the skin, which may make it uncomfortable to wear. Besides, it is unclear how it will work after the user remounts the device or is in motion. EARFace~\cite{zhang2023i} achieves promising facial expression tracking performance on earphones powered by acoustic sensing. However, their system needs to operate in a frequency range as high as $40 kHz$, requiring a sampling rate of at least $80 kHz$, which cannot be satisfied by many commodity speakers, microphones and audio interfaces. Moreover, EARFace emits acoustic signals into the ears and depends on the reflections from the ear canals to track facial expressions. It is not clear whether their system can work on glasses as well as on earphones since the sensing area is completely different and the signal reflection is more complicated outside human body.

In order to compare our work to EarIO, which also uses acoustic sensing~\cite{li2022eario} to infer facial expressions from the movements on the back of the chin using two earables, we conducted a rigorous and thorough comparison and demonstrated that our EyeEcho sensing system outperforms EarIO in terms of the performance, the training data needed, the ability to detect blinks, and stability. We detail our investigation into this matter in Sec.~\ref{Subsec: comparision with eario}.

\section{Background}
\label{Sec: background}
In this section, we introduce the background information on the following aspects: (1) a definition of continuous facial expression tracking; (2) principle of operation of EyeEcho.

\subsection{Continuous Facial Expression Tracking}

In order to position our work better in the previous literature, we would like to provide a precise definition of continuous facial expression tracking. Generally, there are two types of systems for monitoring facial expressions, determined by whether the task they aim to solve is classification task or regression task:

The first category focuses on recognizing a set of pre-defined discrete facial expressions or gestures. Most prior wearable sensing systems \cite{xie2021acoustic,verma2021e,ando2017canalsense,iravantchi2019interferi} fall into this category. They perform classification tasks and report the accuracy of distinguishing these facial expressions or gestures as performance metrics. However, the output of these systems does not provide information on how the face appears during the process of making a facial expression. This information is needed for downstream applications such as adding facial expressions to render a personalized avatar or enabling video conferencing on smart glasses without the need of holding a camera in front of the face.

The second category aims to estimate the position and shapes of all parts of the face (such as the nose, eyes, eyebrows, cheeks, and mouth) continuously, often multiple times per second. Technologies in this category usually carry on regression tasks. Most camera-based methods have been able to track facial expressions continuously. However, achieving continuous tracking with wearable sensing systems has been challenging, as it requires high-quality and reliable information about facial movements. Recently, several wearable methods have started exploring continuous estimation of facial expressions \cite{chen2020c,chen2021n,wu2021bioface,li2022eario,kim2023pan,zhang2023i}. As a starting point, they define a set of facial expressions for participants to perform. Unlike the methods in the first category, these systems are able to continuously provide the position and shapes of different facial parts as a user makes a facial expression, ranging from a neutral face to the most extreme state. The position and shapes of the facial expressions are represented using landmark parameters \cite{chen2020c,wu2021bioface,kim2023pan,zhang2023i} or blendshape parameters \cite{chen2021n,li2022eario}, which are the output of these sensing systems. The performance is measured using the Mean Absolute Error (MAE) between the predicted parameters and the ground truth captured by a frontal camera. EyeEcho belongs to the second category, as it tracks facial expressions continuously using wearables. In Subsection \ref{Subsec: applications}, we further discuss various potential applications that can be enabled by the ability to continuously track facial expressions on glasses.

\subsection{Principle of Operation}

Prior work~\cite{li2022eario,wu2021bioface,zhang2023i,chen2020c} have demonstrated that partial skin and muscle deformations behind and inside ears are highly informative to reconstruct full facial expressions when they are captured by different kinds of sensors. Xie et al.~\cite{xie2021acoustic} proved that the skin and muscle deformations around the eyes and the cheeks contain information that can be extracted to recognize upper facial gestures.
The sensing hypothesis of EyeEcho is that these deformations around eyes and cheeks are highly informative about detailed facial movements on both lower and upper face including eyes, eyebrows, cheeks, and mouth. Considering that people's facial skin and muscles are interconnected, moving any part of the face would inevitably stretch the muscles and skin on the entire face. 
EyeEcho applies this \textit{skin-deformation-based sensing hypothesis} on glasses to develop research questions: \textit{is it possible to infer full facial expressions by only observing skin deformations around glasses (e.g., cheeks)?} If so, what is the appropriate hardware set up including the number, orientation and position of the sensors? How well can it track facial movements on different areas of the face (e.g., blinking) under different real-world scenarios? To explore these research questions, we developed EyeEcho using active acoustic sensing to capture the skin deformations on glasses, which we will detail later. We chose acoustic sensing because its sensors are small, lightweight, low-power and have been successfully applied to various tasks on tracking human activities, including health-related activities detection \cite{nandakumar2015contactless,wang2018c}, novel interaction methods \cite{aumi2013doplink,wang2016device,yun2017strata,xu2020earbuddy}, silent speech recognition~\cite{zhang2020endophasia,zhang2023echospeech,zhang2023hpspeech,sun2023echonose}, authentication~\cite{lu2019lip,gao2019earecho,wang2022toothsonic,Huh2023wristacoustic,iwakiri2023user}, discrete facial expression recognition~\cite{xie2021acoustic}, gaze tracking~\cite{li2024gazetrak}, finger tracking~\cite{sun2018vskin,nandakumar2016fingerio}, hand gesture recognition~\cite{zhang2018fingerping,lee2024echowrist}, body pose estimation~\cite{mahmud2023posesonic}, and motion tracking~\cite{zhang2018vernier,lian2021echospot}.
\section{Design and Implementation of EyeEcho}
\label{Sec: Design}

\subsection{Hardware Prototype Design}
\label{Subsec: form factor}

\begin{figure*}[t]
    \centering
    \subfloat[Customized PCB]{
        \includegraphics[height=.18\textwidth]{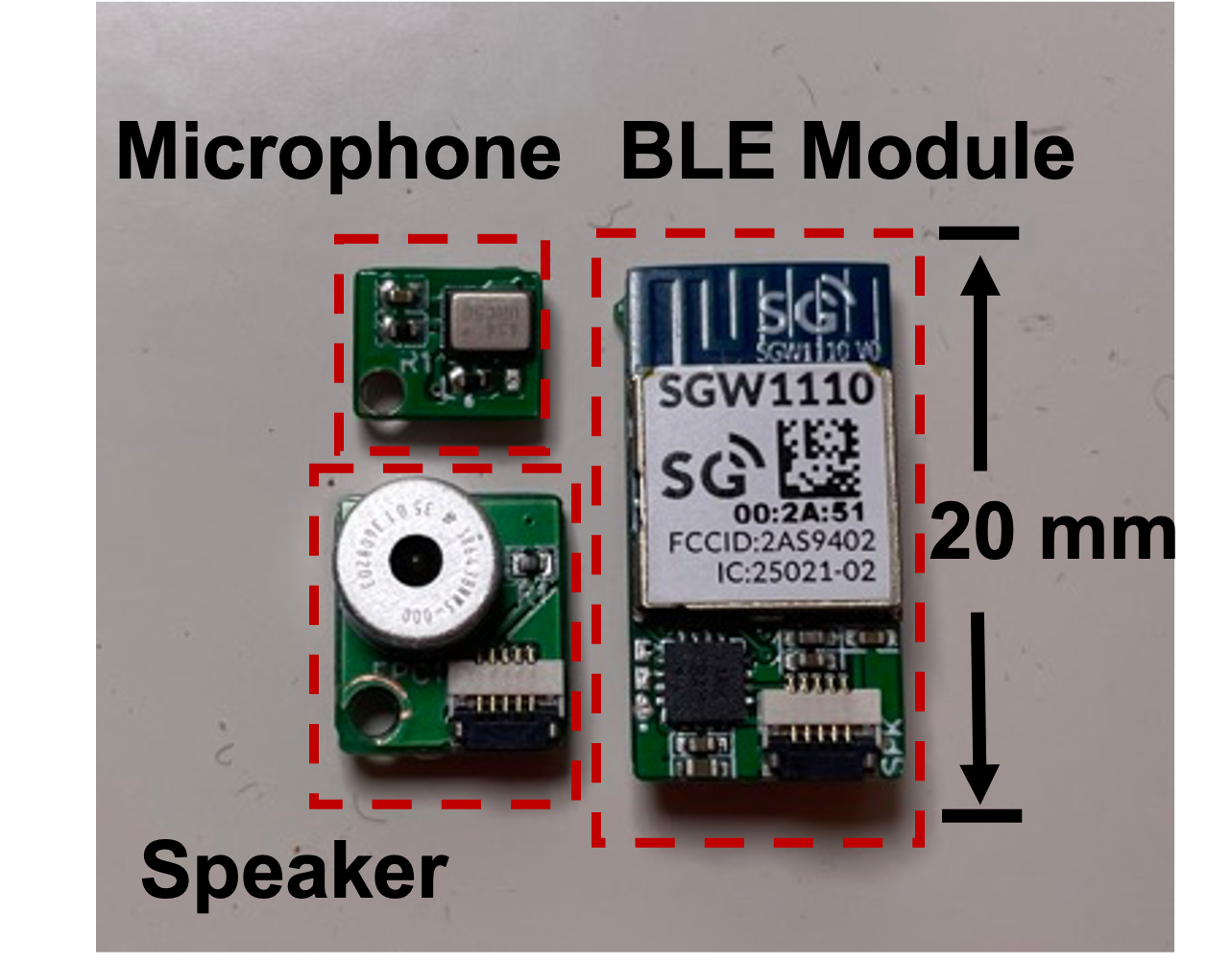}
    }
    \subfloat[Different Sensor Positions]{
        \includegraphics[height=.18\textwidth]{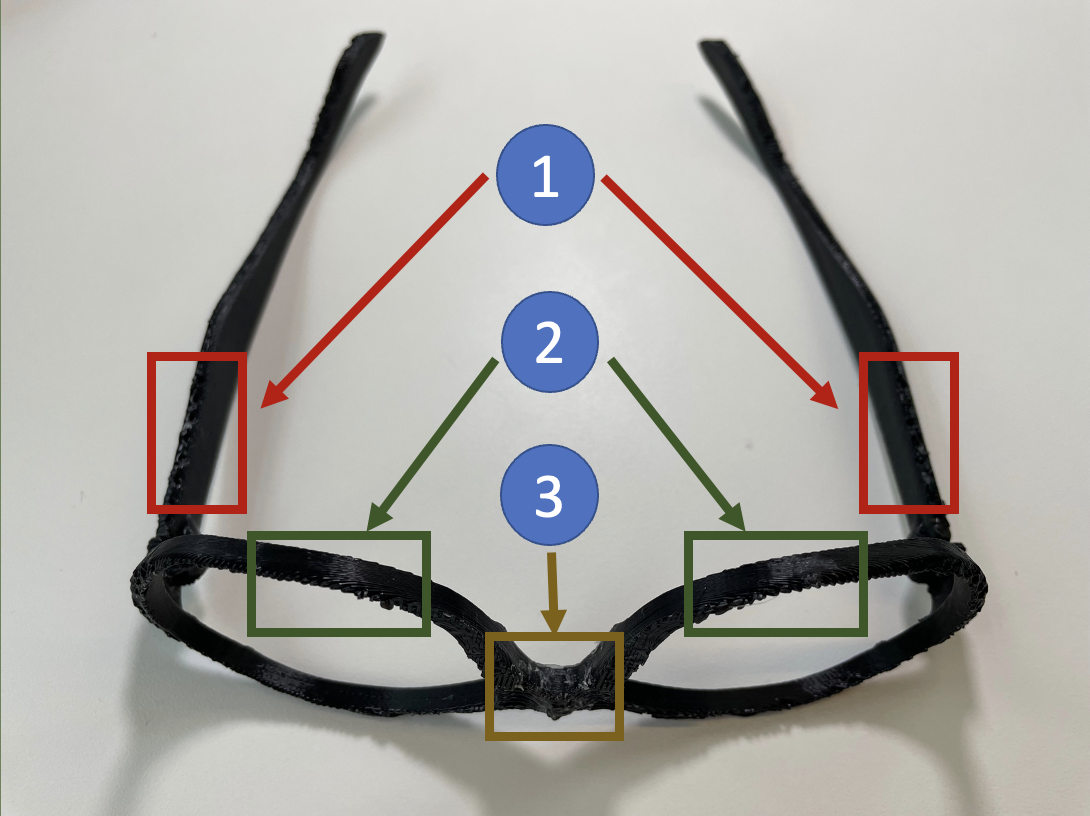}
    }
    \subfloat[Final Prototype]{
        \includegraphics[height=.18\textwidth]{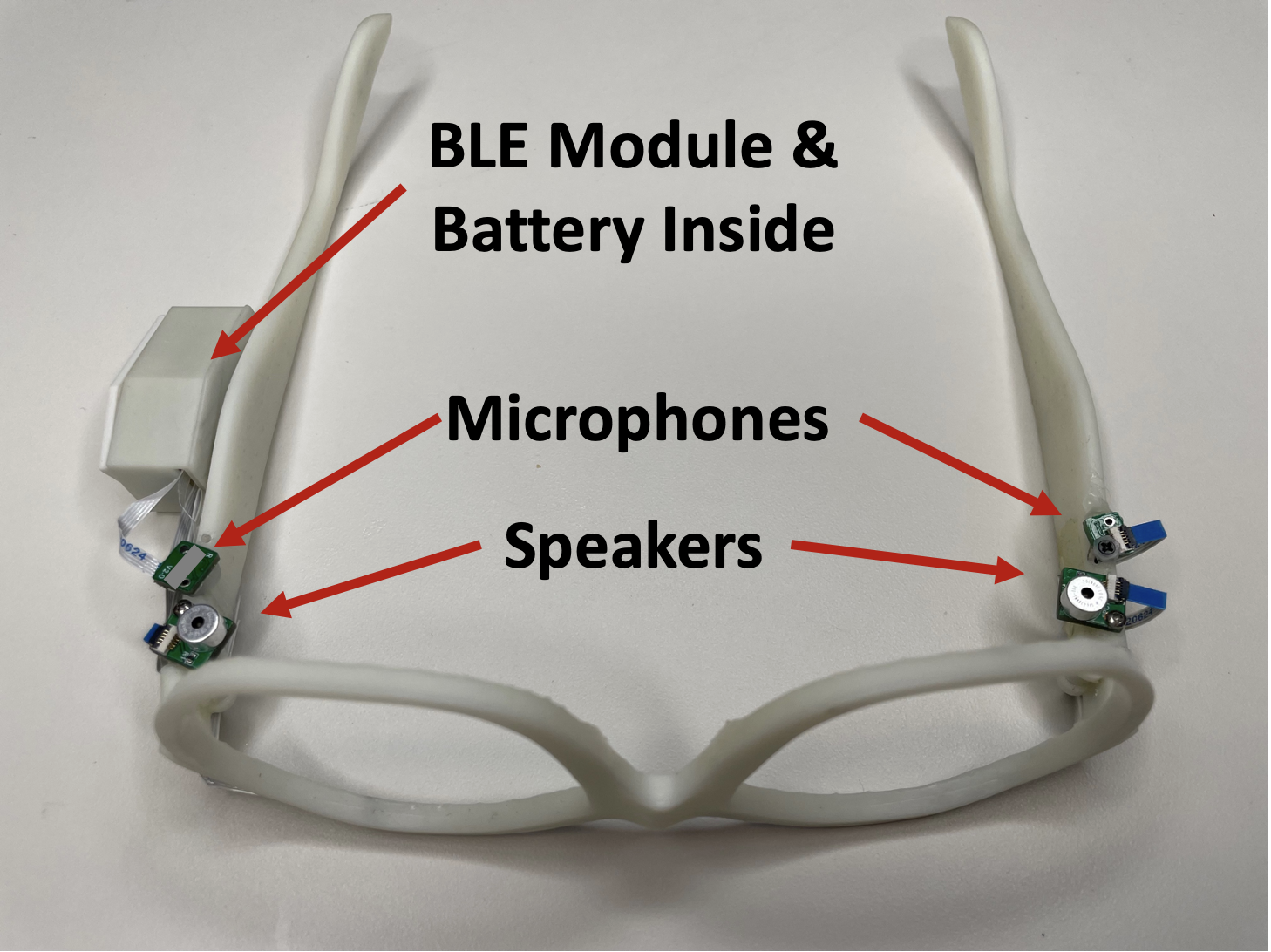}
    }
    \subfloat[Wear the Prototype]{
        \includegraphics[height=.18\textwidth]{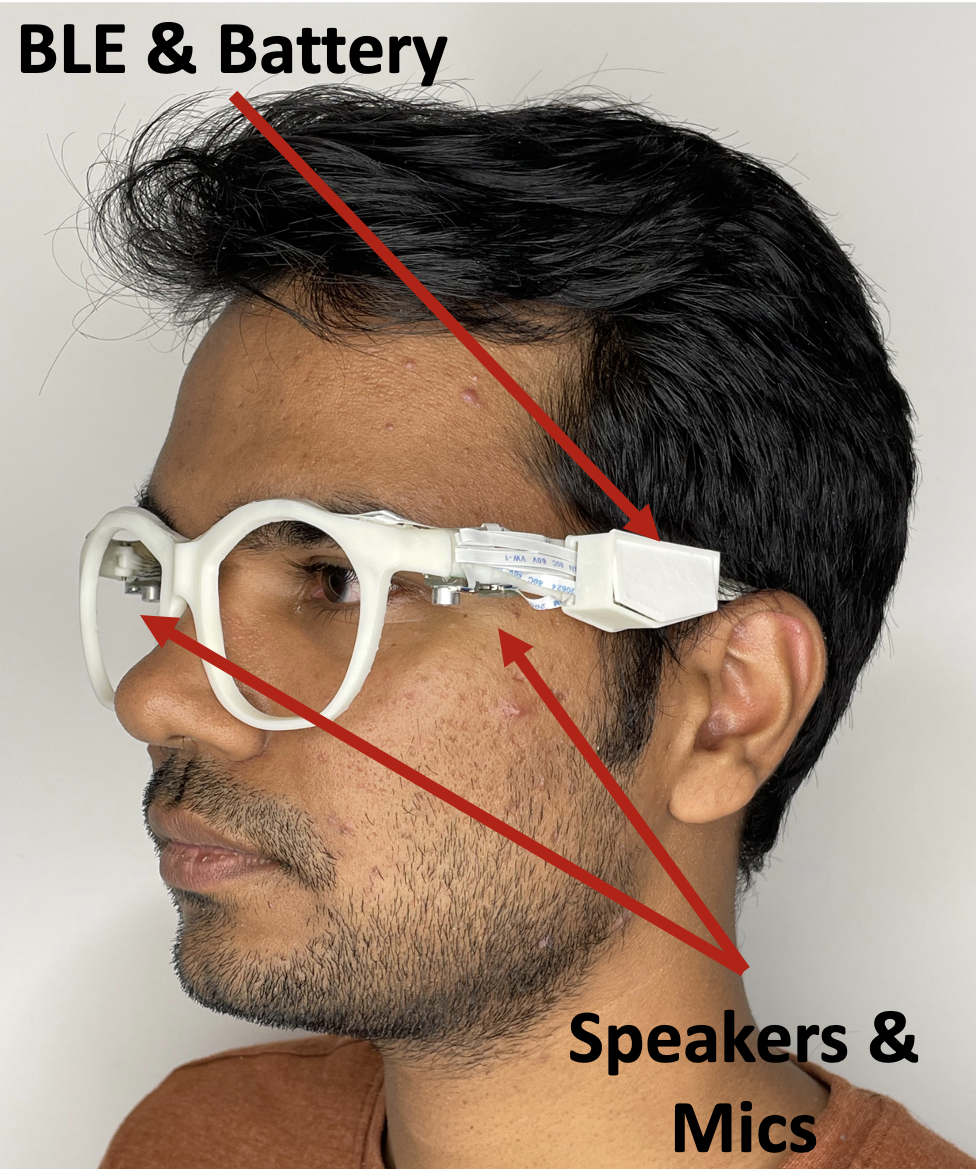}
    }
    \caption{Design of Hardware and Glasses Form Factor.}
    \label{Fig: Design}
    \Description{This figure features four subfigures. Subfigure (a) shows the speakers, microphones and the BLE module on our customized PCBs, which are used in our system. The BLE module is 20 mm in length. Subfigure (b) displays a pair of glasses and it is marked with three positions, i.e., Position 1 on the legs of the glasses, Position 2 on the bottom of the glass frame and Position 3 is under the bridge of the glass frame. Subfigure (c) shows the complete prototype of the glasses embedded with our EyeEcho system. A pair of speaker and microphone is placed on each leg of the glasses. A 3D printed case is placed on the left leg of the glasses housing the BLE module and the battery. Subfigure (d) shows a person wearing the glasses prototype, demonstrating the position of the speakers, microphones, and the case relative to the user's face.}
\end{figure*}

\subsubsection{MCU and Sensor Selection}
The core sensing hardware of our acoustic-based tracking system includes two MEMS microphones (ICS-43434~\cite{ics43434}), two speakers (SR6438NWS-000~\cite{sr6438}) and a bluetooth module (SGW1110~\cite{sgw1110}) housed on customized Printed Circuit Boards (PCB), as shown in Fig.~\ref{Fig: Design} (a). The Nordic's nRF52840 Bluetooth Low Energy (BLE) 5.0 System-on-Chip~(SoC)~\cite{nRF52840} in the Bluetooth module drives the speaker, microphone and data transmissions. The speakers and microphones are both connected to the Inter-IC Sound~(I2S) interface of nRF52840 via FPC wires. We set the sample depth of the data as $8 bit$, which requires a bit rate of about $800 kbps$ to transmit two channels of received data to a smartphone (Xiaomi Redmi) via BLE in real-time without significant packet loss. This is well supported by BLE 5.0~\cite{blethroughput}. This setup of the core acoustic sensing unit allows transmitting two channels of acoustic data reliably via BLE without compromising the performance.

\subsubsection{Exploring Different Sensor Positions on Glasses}
To implement EyeEcho on the form factor of glasses (e.g., smart glasses, AR glasses), we fabricated a pair of glasses using 3D printing and incorporated multiple holes at various positions. These openings allowed us to securely attach the speakers and microphones using screws. Given the limited space available on glasses, we identified three primary positions suitable for sensor placement to capture skin deformations without interfering with users' daily activities: 1) on the legs of the glasses; 2) at the bottom of the frame; 3) under the bridge of the frame, as depicted in Fig.~\ref{Fig: Design} (b).

Position 1 involves placing one pair of speakers and microphones on each leg of the glasses, directed downward toward the user's face. This setup enables the capture of skin deformations on both the left and right sides of the cheeks, facilitating the detection of respective facial movements on each side. The second option, Position 2, relocates the two pairs of speakers and microphones to the bottom of the glasses frame, also pointing downward and closer to the user's cheeks. In addition to these two setups, the last option involves positioning only one speaker under the bridge of the glasses frame (Position 3), while keeping the microphones at Position 2. This configuration consumes less energy as it requires only one speaker.

We conducted a preliminary cross-session experiment involving two researchers to assess the tracking performance of the three positions using the algorithms detailed in the subsequent sections. Across the three setups, we obtained average Mean Absolute Errors (MAE) of 18.0, 18.0, and 22.8, respectively, for 52 blendshape parameters when comparing the ground truth with our predictions. Based on these experimental results, the first two setups outperformed the last option in terms of tracking performance. Notably, Position 2 was found to be more obtrusive than Position 1, as the sensing unit on the glasses frame had a higher likelihood of obstructing the user's face and impacting their daily activities. Furthermore, it's worth noting that most commodity smart glasses integrate sensors on their legs. Therefore, placing the sensors on the legs aligns with the potential future adoption of this sensing technology on glasses. Consequently, we chose Position 1, which involves attaching a pair of speakers and microphones to each leg of the glasses.

\subsubsection{Form Factor Design}
After finalizing the positions of the speakers and microphones, we designed and 3D printed one case with a sliding cover to house the BLE module and the battery. The case is mounted on the left leg of the glasses and specifically designed to match the shape of the leg. We cut a narrow slot on one side of the case in order to let the FPC wires go through and be connected with the speakers and microphones. While connecting the speakers and microphones to the BLE module, the FPC wires are routed along the glass frame, so that the wires will not block the user's view and be less obtrusive. The final complete prototype is shown in Fig.~\ref{Fig: Design} (c). We think the final prototype is minimally-obtrusive and very close to a normal glass frame in appearance, as shown in Fig.~\ref{Fig: Design} (d).

\begin{figure*}[t]
  \includegraphics[width=1\textwidth]{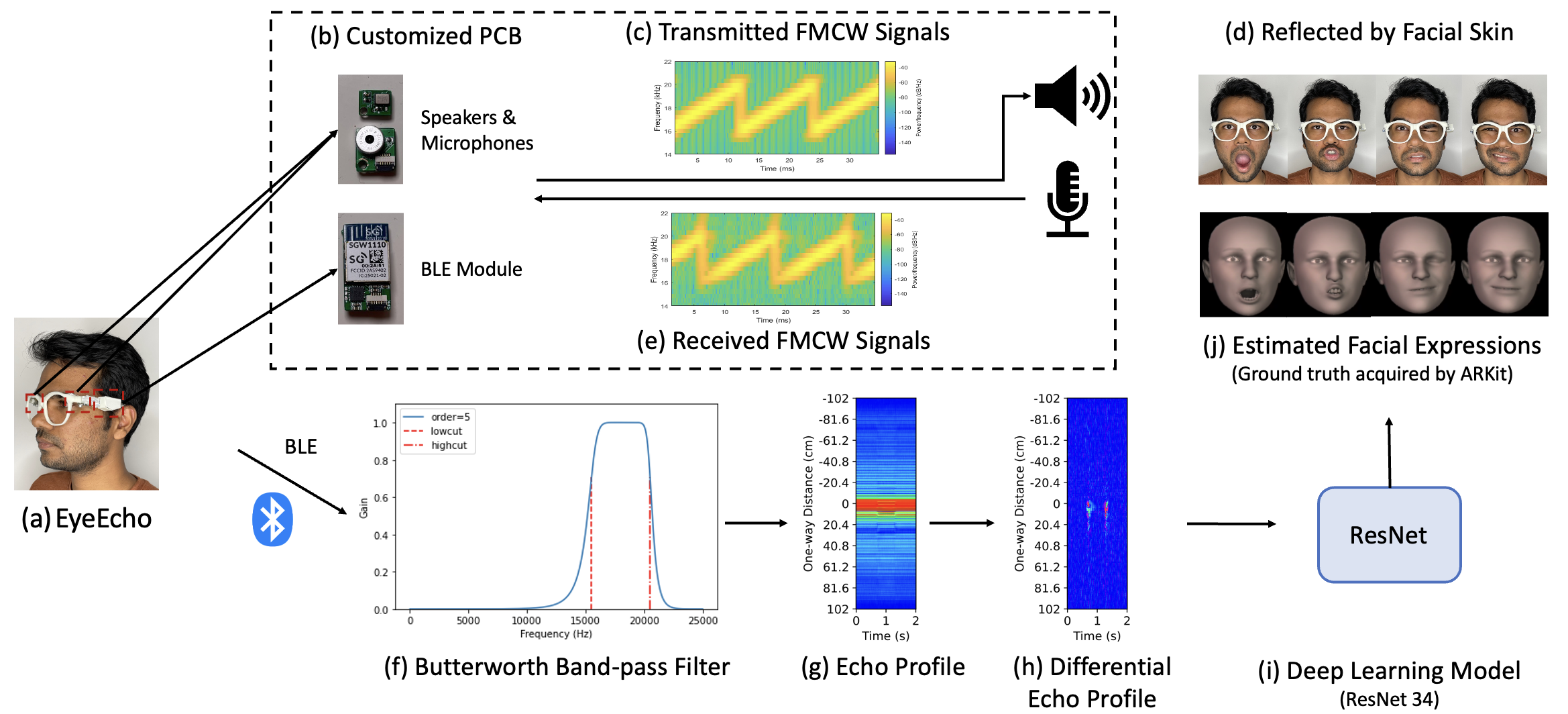}
  \caption{Overview of EyeEcho System.}
  \label{Fig: system overview}
  \Description{This figure displays the overview of the EyeEcho system. Subfigure (a) shows a person wearing the glasses prototype. Subfigure (b) shows the PCBs of speakers, microphones and BLE module. Subfigure (c) shows the transmitted FMCW signal in time-frequency domain. This signal sweeps from 16-20 kHz. Subfigure (d) shows the signal reflected from the facial skin. Subfigure (e) shows the received FMCW signal in time-frequency domain. The signal sweeps from 16-20 kHz but has some noises in it. Subfigure (f) shows a Butterworth Band-pass filter from 16-20 kHz in the frequency domain. Subfigure (g) shows the echo profile of one facial expression. It displays the unique patterns associated with the facial expression. Subfigure (h) shows the differential echo profile which more clearly demonstrates the pattern of the facial expression. Subfigure (i) shows a deep learning module with the ResNet-34 architecture. Subfigure (j) shows the avatar with the estimated facial expressions corresponding to the facial expressions that the person performs.}
\end{figure*}

\subsection{Sensing Skin Deformations using FMCW-based Acoustic Sensing}
\label{Subsec: acoustic sensing}
With the prototype of glasses completed, we then introduce how we adopt FMCW-based acoustic sensing to sense the skin deformations on glasses. The system overview of EyeEcho is displayed in Fig.~\ref{Fig: system overview}.

\subsubsection{FMCW Signal Transmission}

We choose FMCW as the acoustic sensing technique, because it has demonstrated robust performance on wearable-based sensing applications, including tracking finger positions~\cite{nandakumar2016fingerio}, breathing patterns~\cite{nandakumar2015contactless,wang2018c}, and skin deformations around ear \cite{li2022eario}. We set the frequency range of transmitted FMCW signals in our system to $16-20 kHz$, which is nearly inaudible to most adults and can be reliably achieved by most commodity speakers and microphones as demonstrated in many prior work \cite{nandakumar2016fingerio,zhang2020endophasia}. In order to emit the transmitted signals in this ultrasonic frequency range, we set the sampling rate at $50 kHz$, which can support the frequency range up to $25 kHz$ in theory. Besides, the FMCW sample length is set as 600, which is 0.012 seconds long ($600~samples / 50 kHz$). In this setting, our system is capable of estimating facial expressions 83.3 times per second, which is enough to proivde similar sample rate to as video recordings (30 or 60 frames per second~(FPS)). This FMCW signal is pre-generated as shown in Fig.~\ref{Fig: system overview} (c), and stored in the BLE module to drive the speakers.

\subsubsection{Echo Profile Calculation}
Once acoustic signals are reflected by the face and received by the microphone (Fig.~\ref{Fig: system overview} (e)), we conduct further data processing. We first apply a 5-order Butterworth band-pass filter with low-cut and high-cut frequencies as $15.5 kHz$ and $20.5 kHz$ to filter out noise that is outside the frequency range of our interest, as shown in Fig.~\ref{Fig: system overview} (f). Then we obtain the \textit{Echo Profile} of received signals by calculating the cross-correlation between the received signals and transmitted signals~\cite{wang2018c}, which can display the deformations of the skin in both temporal and spacial domains (Fig.~\ref{Fig: system overview} (g)). 

With the echo profiles calculated, EyeEcho provides a spacial tracking resolution of $6.8 mm$ ($340 m/s$ / $50 kHz$) and a maximum tracking distance of $4.08 m$ ($6.8 mm \times 600~samples$). Because the speaker and microphone are placed very close to each other, the signals travel in round trips from the speaker to the face and back to the microphone. Thus, the real tracking resolution in space and the maximum one-way tracking distance are $3.4 mm$ and $2.04 m$ respectively. This means that each "echo bin" in the echo profile reports the total signal strength our system receives at a specific distance from our system and two consecutive echo bins are $3.4 mm$ apart from each other. As the skin deforms by small amount, the way the signal strength is distributed in different bins is changing. Observing these changes in the echo profile makes the EyeEcho system work. In order to remove the static objects in the background and alleviate the impacts of remounting the device, we further calculated \textit{Differential Echo Profile} by subtracting the echo profile between two adjacent echo frames, like the one shown in Fig.~\ref{Fig: system overview} (h). Please note that signals at the negative distance in the echo profile are the reflected signals of the last echo frame from a very long distance and are usually useless in tracking facial expressions. 

\begin{figure*}[t]
  \includegraphics[width=1 \textwidth]{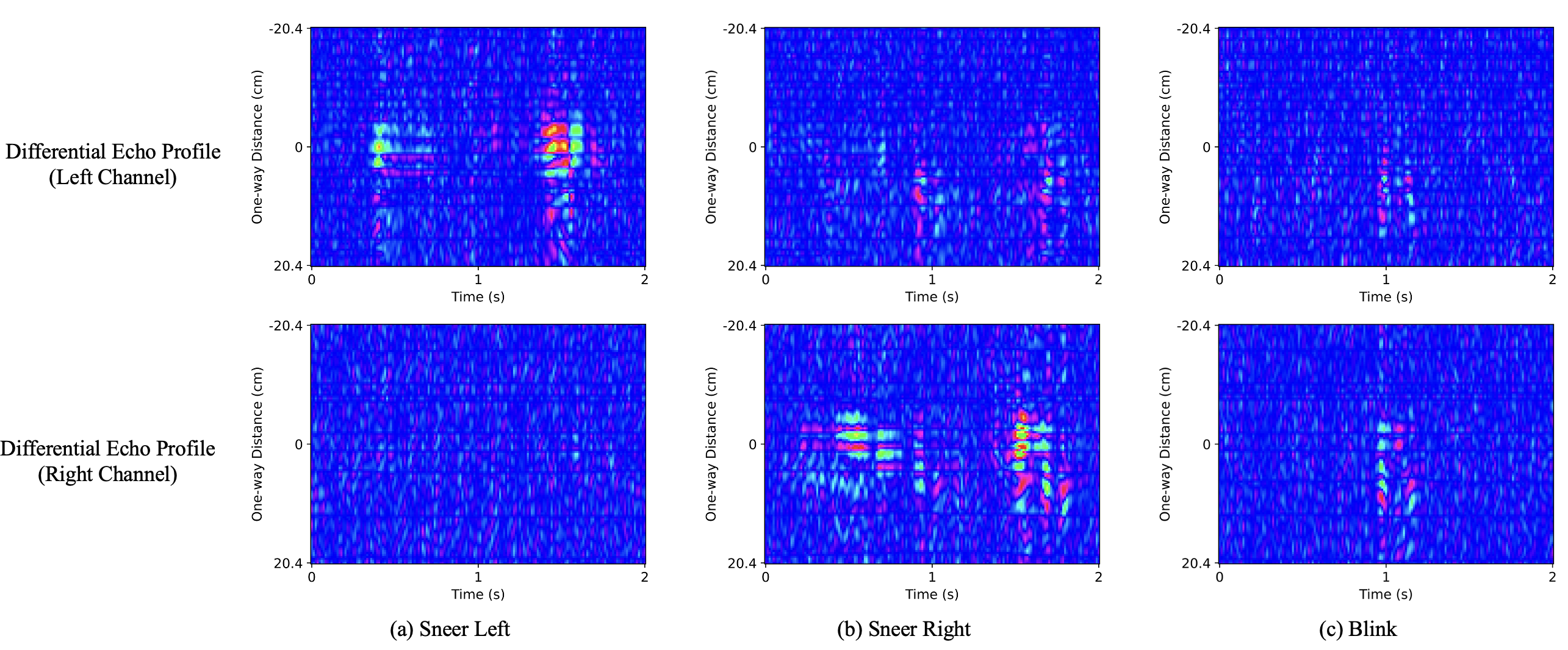}
  \caption{Comparison of Differential Echo Profiles of Three Facial Movements.}
  \label{Fig: comparison of patterns}
  \Description{This figure shows the differential echo profiles of three facial movements in both left channel and right channel. For Sneer Left and Sneer Right, there is a clear pattern in the corresponding channel in the differential echo profiles. There is a weaker pattern in the other channel. For subtle facial movement like blinking, the pattern is weaker and shorter which aligns with nature of this movement.}
\end{figure*}

Fig.~\ref{Fig: comparison of patterns} demonstrates the differential echo profiles of three facial movements performed by a researcher. When the user performs one facial expression that only relates to one half of the face (Fig.~\ref{Fig: comparison of patterns} (a) and (b)), a clear pattern can be observed in the differential echo profile in the corresponding channel. The other channel also contains weaker information because people's facial skin and muscles are interconnected. Moreover, the major parts of the patterns are within $0-10 cm$ because the user's face is located within this distance from our EyeEcho system approximately. Because the signal we emit is finite in the frequency domain, it diffuses into further echo bins in the time domain when the cross-correlation is calculated. Also considering the multi-path reflection, some patterns can be observed at the negative distance and the distance beyond $10 cm$. However, the echo bins between $0-10 cm$ contain most information for EyeEcho to track facial expressions. When the user blinks (Fig.~\ref{Fig: comparison of patterns} (c)), patterns can be observed in the differential echo profiles of both channels. Compared with the patterns caused by sneering left or right, the patterns related to blinking are weaker and shorter because of the nature of people's blinks. All these features support our hypothesis that the skin and muscle deformations around eyes and cheeks are informative for both upper-face and lower-face movements and can be captured by acoustic sensors to track users' full facial expressions.

Since the distance between the glasses and the face is usually under $10 cm$, we remove any echoes that are beyond $10.2 cm$ ($3.4 mm \times 30~echo~bins$) in the differential echo profile. This will help us to minimize the impact of the acoustic echos from environment objects which are usually placed at a much further distance. This differential echo profile with a length of one second is sent to a customized learning algorithm to estimate the full facial expressions, as detailed in Sec.~\ref{Subsec: learning algorithm}.

\subsection {Learning Algorithms for Continuous Facial Expression Tracking}
\label{Subsec: learning algorithm}
In order to infer full facial expressions continuously from the skin deformations represented by differential echo profiles, we adopted a customized deep learning pipeline.  

\subsubsection{Ground Truth Acquisition}
\label{Subsubsec: ground truth}
The deep learning model requires reliable ground truth of the facial expressions to train the model. We choose to use the TrueDepth camera on an iPhone powered by Apple's ARKit API~\cite{arkit} to record the ground truth of facial expressions at 30 FPS, represented by 52 blendshape parameters. Each parameter is in charge of the shape and position of one part of the face (e.g., jawOpen). The original range of each blendshape parameter is from 0 to 1. We multiply the value of each blendshape parameter by 1000 to better train the model. After this operation, the maximum possible value of each parameter is 1000. As demonstrated in previous work~\cite{chen2021n,li2022eario}, this method can provide reliably track 3D facial movements on the cheeks, the eyes, the eyebrows, the nose, the mouth, and the tongue. We chose to use this blendshape-based ground truth instead of 2D landmarks because it can provide more visual expressiveness by showing the movements of different parts of the face in 3D.

\subsubsection{Deep Learning Model}
\label{Subsec: Learning Model}

We apply a sliding window of one second on the received acoustic data, which leads to 84 frames in Differential Echo Profiles of each window. Each Differential Echo Profile has 60 data points including 30 data points representing the echo distance of $10.2 cm$ for each microphone. In total, the dimension of the input vector for the deep learning model is $60 \times 84$. To make predictions for the current frame, we utilize the current frame plus the 83 frames prior to the current frame as input data so that there is minimum delay for the real-time prediction. Technically, the maximum possible delay caused by the sliding window is the length of one frame which is $12 ms$. As we apply the sliding window technique, for each prediction, we can store and reuse the last 83 echo frames from last prediction and just calculate the current echo frame so that the calculation will be fast enough for real-time prediction at our expected refresh rate.

To estimate full facial expressions from these Differential Echo Profiles, we adopted an end-to-end Convolutional Neural Network (CNN) model. We chose CNN in our system rather than other models such as Recurrent Neural Network (RNN) because after processing the raw audio data and obtaining the Differential Echo Profile, the time series data is converted to images containing information in both spatial and temporal domains. Adopting CNN achieves a better performance in this kind of tasks. Thus, a 34-layer Residual Neural Network (ResNet-34) is used as the backbone learn the features from the input data vectors and a fully-connected decoder is utilized to estimate the facial expressions including 52 parameters of the blendshapes. MAE between the prediction and the ground truth was used as the evaluation metric to train the model.

\section{Evaluation of EyeEcho in an In-lab Study}\label{Sec: user study}

\begin{figure*}[t]
  \includegraphics[width=0.8\textwidth]{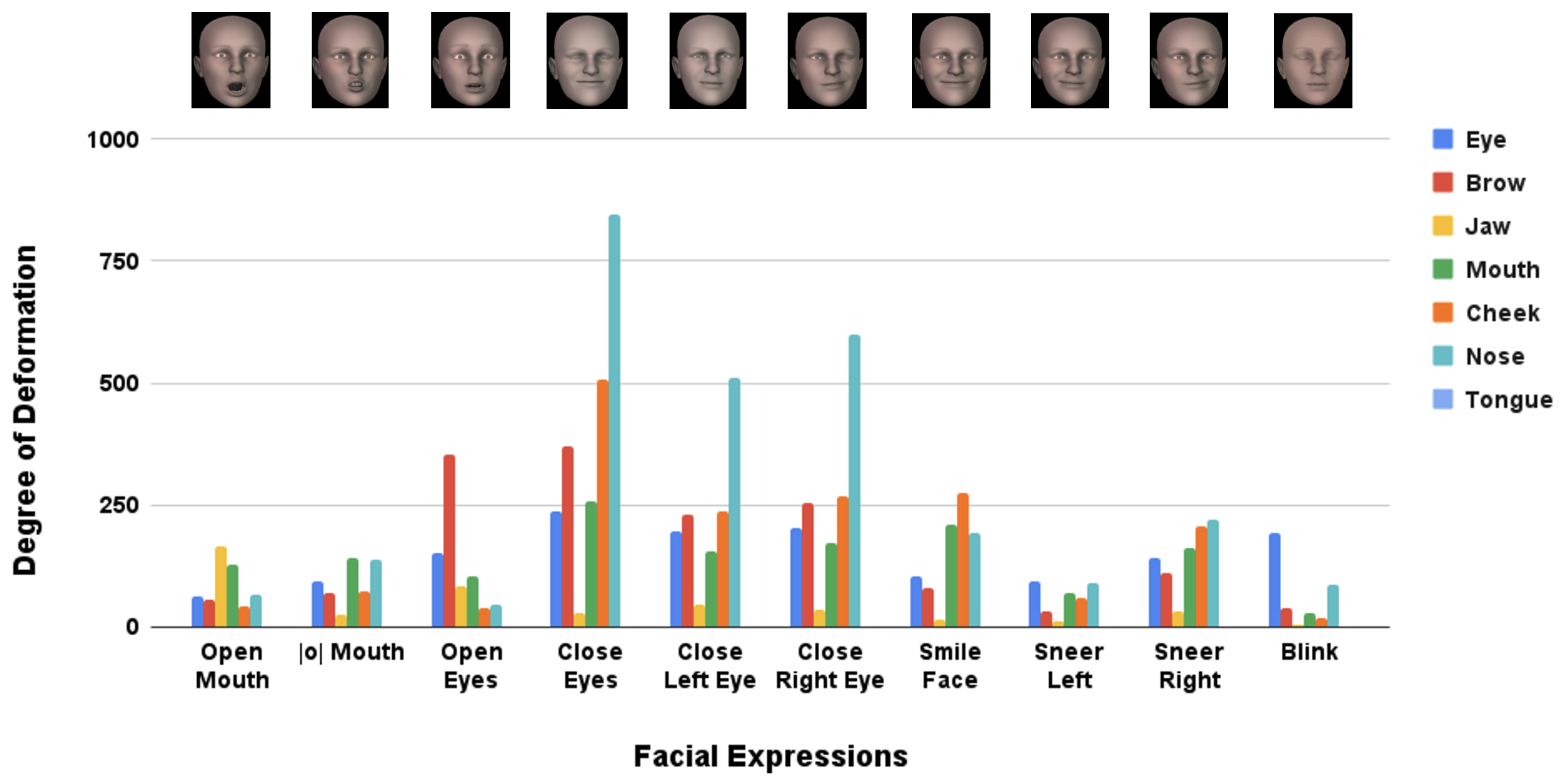}
  \caption{The Degree of Deformation of Different Facial Parts when Performing Different Facial Expressions to the Most Extreme State.}
  \label{Fig: expressions with strength}
  \Description{This figure shows the degree of deformation of different facial parts, including eye, eyebrow, jaw, mouth, cheek, nose and tongue, when the nine facial expressions and blinking are performed. When any of the facial expressions is performed, all facial parts deform to a certain extent for both lower and upper facial parts.}
\end{figure*}

We aim to design a facial expression tracking system that is light-weight, low-power and robust in various real-life scenarios. To achieve this, we evaluated EyeEcho with both controlled in-lab study and naturalistic semi-in-the-wild study. The in-lab study aims to provide an in-depth analysis of the performance of EyeEcho with different controlled setups, more granular metrics, and more comprehensive experiments, while the semi-in-the-wild study focuses on showcasing EyeEcho's actual performance in near-real-world settings.
We present the details of the first in-lab study in this section followed by Sec.~\ref{Sec: evaluation with different settings} which introduces another in-lab study evaluating EyeEcho comparing different settings, while details of the semi-in-the-wild study are in Sec. \ref{Subsec: wild study}.

In the first study, we considered several scenarios that commonly occur in real life, such as sitting, walking, and remounting. The validation within these contexts serves as a crucial step in establishing the system's potential effectiveness in real-world applications.

\subsection{Study Design}\label{Subsec: study design}
\subsubsection{Apparatus}
We used the hardware and form factor described in Sec.~\ref{Subsec: form factor} to conduct the study. We used a smartphone, Xiaomi Redmi to function as a server, receiving transmitted data from the BLE module. To record the ground truth of facial movements and play the instruction video, an iPhone 12 with the TrueDepth camera was placed in front of participants. In walking scenario, participants wore a chest mount which kept the iPhone in front of the face.

\subsubsection{Selection of Facial Expressions for Continuous Tracking}
\label{Subsec: facial expressions}
We selected nine distinct facial expressions that involve movements of both the upper and lower face, as visually represented in Fig.~\ref{Fig: Facial Expressions with Different Echo Profiles}. It's important to note that while these expressions were chosen for testing purposes, our system continuously tracks facial expressions. These particular expressions were carefully chosen to enable a comprehensive evaluation of EyeEcho's performance in estimating a wide range of common facial expressions, encompassing movements in various facial regions, including the eyes, eyebrows, mouth, and cheeks.

To emphasize this point, we initially defined the "degree of deformation of one facial part" as the average of the blendshape parameters associated with that specific part out of the 52 blendshape parameters provided by the ARKit API, as introduced in Sec.~\ref{Subsubsec: ground truth}. Subsequently, we plotted the degree of deformation of different facial parts while participants performed these nine facial expressions to the most extreme state, as depicted in Fig.~\ref{Fig: expressions with strength}. As previously discussed in Sec.~\ref{Subsubsec: ground truth}, theoretically, the maximum possible degree of deformation for one facial part is 1000. The figure illustrates that nearly all facial parts, with the exception of the tongue (related to only one blendshape parameter), exhibit some degree of movement when performing either upper or lower facial expressions. This observation underscores the interconnected nature of facial muscles. Moreover, the movements of the mouth, cheeks, and nose are even more pronounced when upper facial expressions, such as "Close Eyes," are performed. This substantiates our claim that these nine selected facial expressions effectively evaluate our system's performance in tracking multiple facial parts simultaneously. It's worth noting that these expressions have also been utilized in prior studies, such as \cite{li2022eario}. During the study, participants were presented with these facial expressions multiple times in an instructional video to mimic, and the order of presentation was randomized in each session to mitigate the influence of expression sequence. In Fig.~\ref{Fig: expressions with strength}, we also included a subtle facial movement, blinks, because they happened spontaneously to the participants in the study.

While the theoretical maximum degree of deformation is 1000, practical facial deformations performed by humans are significantly lower than this maximum value. As illustrated in Fig.~\ref{Fig: expressions with strength}, even when participants reach the most extreme state of a particular facial expression, the degree of deformation for all or most facial parts remains below 250. This observation underscores that in practice, facial deformations are well below the theoretical maximum.

\begin{figure*}[t]
  \includegraphics[width=0.8\textwidth]{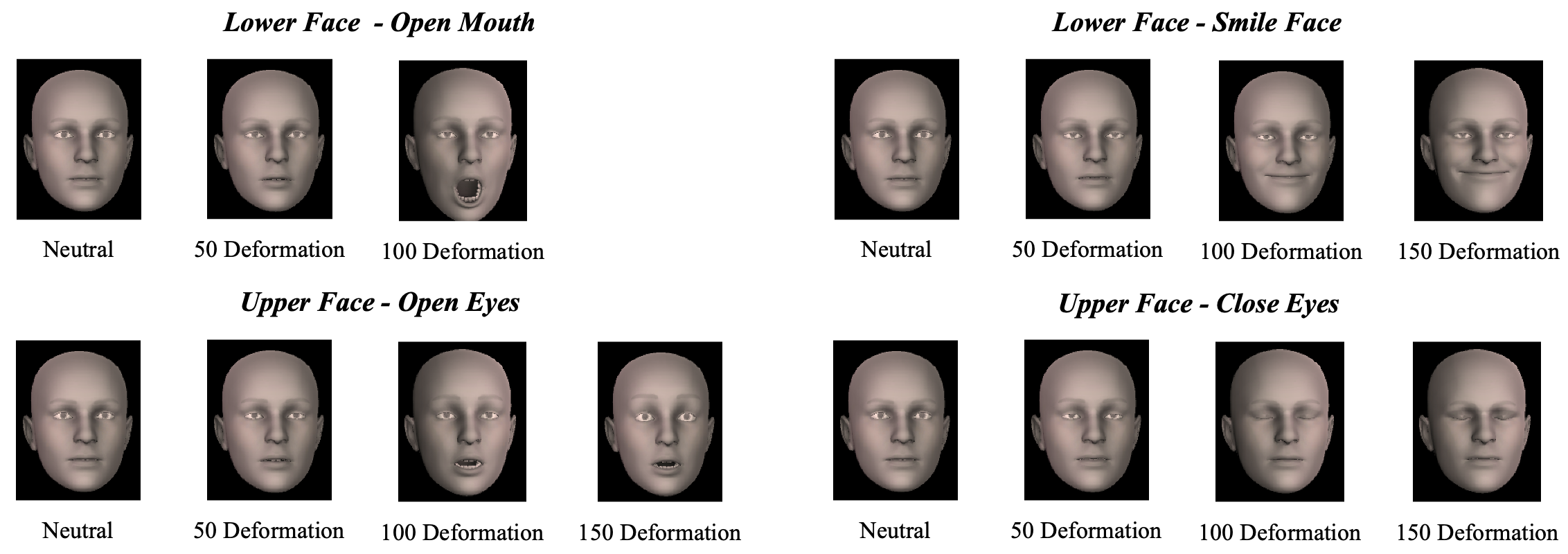}
  \caption{Visualization of Varying Degrees of Deformation.}
  \label{Fig: different deformation}
  \Description{This figure shows the visualization of four facial expressions, Open Mouth, Smile Face, Open Eyes, and Close Eyes, with different degrees of deformation. For Open Mouth, the face reaches the most extreme state with 100 Deformation while for the rest three facial expressions, the face reaches the most extreme state with 150 Deformation.}
\end{figure*}

To provide readers with a visual understanding of varying degrees of deformation for different facial expressions, we calculated the "degree of deformation of one facial expression" by averaging the values of the 52 blendshape parameters representing that specific facial expression. We then visually presented how four different facial expressions appear at different degrees of deformation, specifically when the degree of deformation is 0 (neutral face), 50, 100, and 150, as depicted in Fig.~\ref{Fig: different deformation}. Notably, we could not plot the 150-degree deformation for "Open Mouth" because the degree of deformation could not reach 150 even when the mouth was opened to the extreme. From our empirical observations presented in the figure, it can be inferred that when the degree of deformation reaches 150, most facial expressions visually attain their most extreme state.

\subsubsection{Study Procedure}

This study was approved by the IRB at the researcher's institution. We successfully recruited 12 participants in this study. We have 3 female and 9 male participants, ranging from 18 to 25 years old. Each participant filled out a questionnaire to collect their demographic information and was compensated USD \$15 after they came to participate in the study.

The study was conducted in a large experiment room on campus, across different times of the day. During the user study, participants were asked to remove their own glasses before the study (if applicable) to wear the testing glasses. The evaluation process has two scenarios: sitting and walking. We started the study with the sitting scenario to initially assess the optimal system performance while participants remained relatively stationary. After the sitting scenario, adjustments were made before transitioning to the walking scenario. For example, participants wore a chest mount to hold the iPhone in front of them to capture their faces while walking.

Each scenario comprised 12 sessions, each of which lasted for approximately two minutes. During each of these sessions, the instruction video displayed on the screen of the iPhone placed in front of the participants featured a researcher performing all nine facial expressions six times in a random order, with brief pauses between expressions. Participants were directed to emulate the facial expressions shown in the video. Following the correct placement of the device, participants underwent a two-minute practice session to acquaint themselves with the testing system and the required facial expressions.

Before each session, participants were instructed to clap his/her hands for synchronization between the EyeEcho device and the ground truth acquisition system. After each session, participants were asked to remount the device by themselves, including taking off the device, taking a short break, and putting the device back on. The goal was to evaluate how our system can perform after the device was remounted which introduced shifts on the wearing positions. During the walking scenario, participants were instructed to walk around in the study room at their comfortable walking speed while mimicking the facial expressions displayed on the screen of the iPhone held by the chest mount they wore.

In total, for each participant, we collected around 48 minutes of data from 24 sessions for two scenarios combined. This includes 1296 facial expressions (9 facial expressions $\times$ 6 repetitions $\times$ 12 sessions $\times$ 2 scenarios).
\begin{figure*}[t]
  \includegraphics[width=0.8\textwidth]{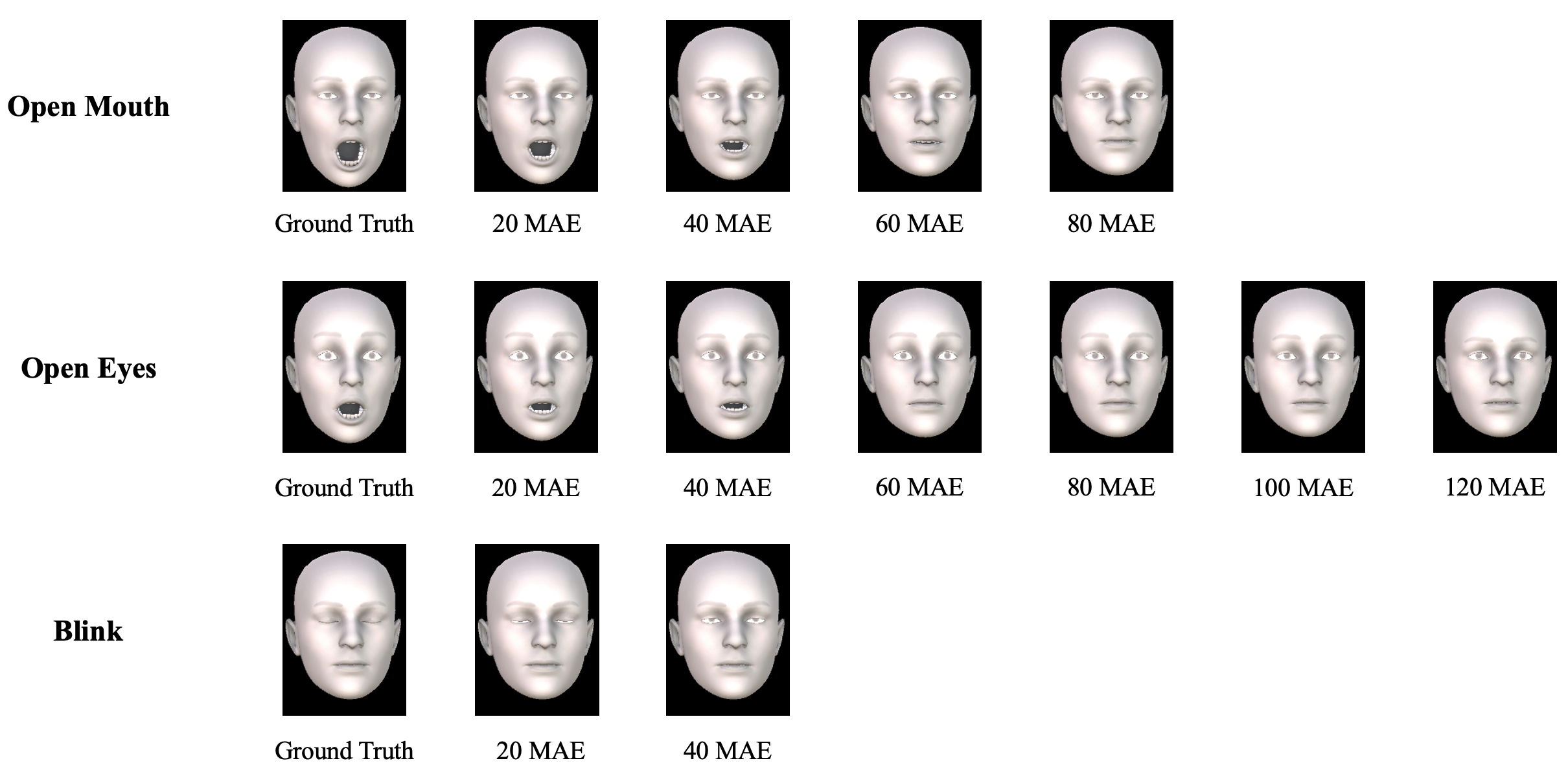}
  \caption{Visualization of Facial Expressions under Different Values of MAE.}
  \label{Fig: Visualized MAE}
  \Description{This figure visualizes the prediction of three facial expressions with different values of MAE compared with ground truth, which are Open Mouth, Open Eyes and Blink. For large facial expressions like Open Mouth and Open Eyes, the prediction is visually similar to ground truth when MAE is 40. When MAE is around 20, the prediction is basically indistinguishable from the ground truth. For subtle facial expressions like blinking, when the MAE is around 20, the prediction is also visually highly similar to the ground truth.}
\end{figure*}

\subsection{Evaluation Metrics}
\label{Subsec: evaluation metrics}
As discussed in Sec.~\ref{Subsubsec: ground truth}, a full facial expression is represented by 52 blendshape parameters in our system. We evaluate the performance using the Mean Absolute Error (MAE) of the 52 parameters between the prediction of EyeEcho and the ground truth. Employing this common metric allows us to compare EyeEcho's performance with that of prior work. It is important to note that the same MAE value can yield substantially different visualization results for the lower face and upper face in terms of how closely the predicted facial expression matches the ground truth, as perceived by human eyes. In order to help readers better understand the true performance of our system, we divided our evaluation metrics into two categories: 1) Lower-face MAE~(LMAE) - evaluating 33 lower-face blendshape parameters related to the movements of cheeks, mouth, nose and tongue and 2) Upper-face MAE~(UMAE) - evaluating the remaining 19 upper-face blendshape parameters related to the movements of eyes and eyebrows. Based on our observation and also from prior work \cite{li2022eario,chen2021n}, there is little visual difference between the prediction and ground truth when LMAE is below 40 and UMAE is below 60. Therefore, we adopted two other evaluation metrics in the results, the Percentage of Frames with LMAE under 40~(PL40) and the Percentage of Frames with UMAE under 60~(PU60). In total, we report five metrics, including MAE, LMAE, UMAE, PL40 and PU60 in the following sections when reporting the tracking performance of EyeEcho.

We plotted the visualization of facial expressions with different MAE for three facial expressions, Open Mouth, Open Eyes, and Blink, in Fig.~\ref{Fig: Visualized MAE}. The figure shows that for large facial movements such as Open Mouth (lower-face) and Open Eyes (upper-face), the prediction is visually similar to the ground truth when MAE is under 40. When MAE is around 20, the prediction is almost indistinguishable to the ground truth. For subtle facial movements, like blinking, the prediction is also highly similar to the ground truth visually when MAE is under 20.

\subsection{User-Dependent Model}
\label{Subsec: user-dependent model}
We first analyze the performance of EyeEcho to track facial expressions continuously. With all the 12 remounting sessions of data collected under each scenario, we conducted a 6-fold cross-validation using 10 sessions to train the model and 2 sessions as the testing sessions to evaluate the results. We report the five evaluation metrics, as shown in Tab.~\ref{Tab: study results}.

\begin{table}[htbp]
\caption{Evaluation Results for both Scenarios.}
\label{Tab: study results}
\begin{tabular}{| c | c | c | c | c | c |} 
\hline
 Scenario & MAE & LMAE & UMAE & PL40 & PU60\\ [0.5ex] 
 \hline\hline
 Sitting & 22.9 & 20.4 & 27.1 & 88.8\% & 92.6\%\\
 \hline
 Walking & 26.9 & 22.7 & 34.3 & 87.1\% & 88.5\%\\
 \hline
\end{tabular}
\end{table}

\begin{figure*}[h]
    \centering
    \subfloat[Sitting Scenario]{
        \includegraphics[height=.3\textwidth]{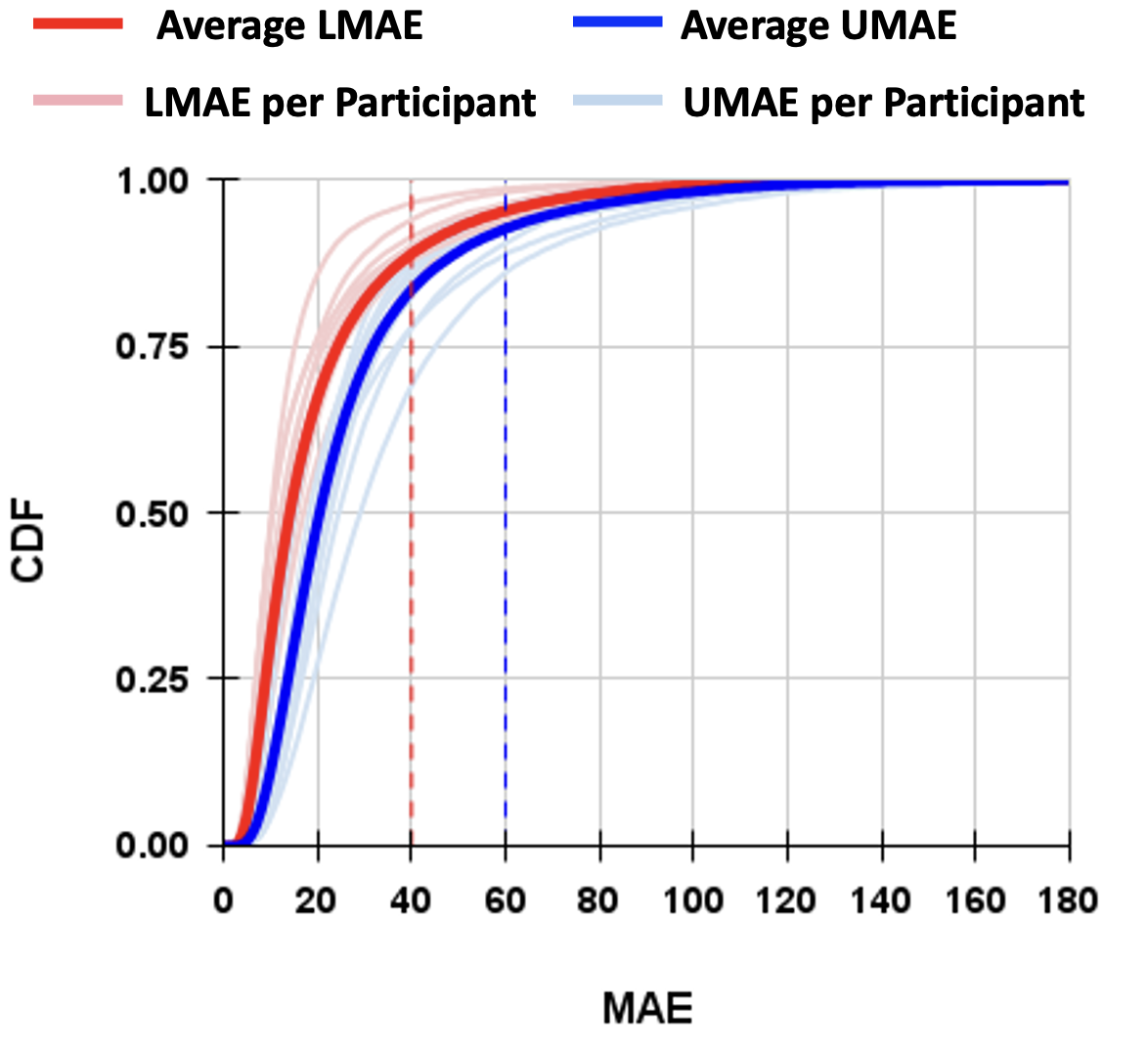}
    }
    \hspace{.1\textwidth}
    \subfloat[Walking Scenario]{
        \includegraphics[height=.3\textwidth]{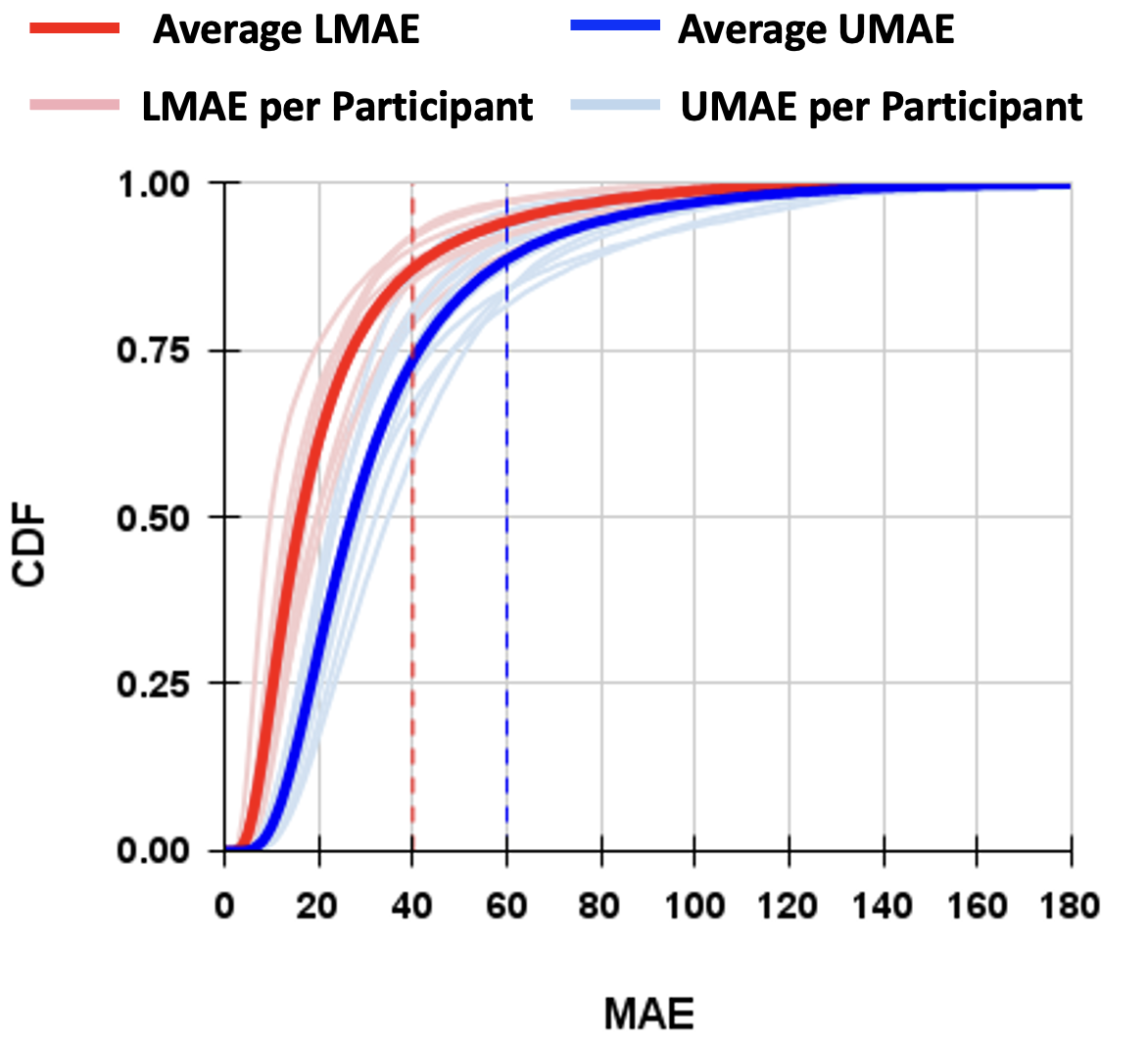}
    }
    \caption{Cumulative Distribution Function~(CDF) of MAE for all Participants. Red Lines: LMAE, Blue Lines: UMAE. Pale Lines: MAE for Each Participant, Solid Lines: Average MAE, Dash Lines: 40 LMAE and 60 UMAE.}
    \label{Fig: cdf}
    \Description{This figure shows the Cumulative Distribution Function (CDF) of MAE for all participants. Subfigure (a) for sitting scenario and Subfigure (b) for walking scenario. Generally, LMAE is smaller than UMAE.}
\end{figure*}

\subsubsection{Numerical Results}
In the sitting scenario, the average MAE for all 12 participants is 22.9, ranging from 16.3 to 27.8 with a standard deviation of 3.2. We plotted the cumulative distribution function~(CDF) of LMAE and UMAE in average and also for each participant in Fig.~\ref{Fig: cdf} (a). These results validate EyeEcho's ability to track the continuous movements of both the lower face and upper face of users in the sitting scenario across different remounting sessions.

In the walking scenario, the average MAE is 26.9, ranging from 20.0 to 31.6 with a standard deviation of 3.1. Same as the sitting scenario, we plotted CDF of LMAE and UMAE in Fig.~\ref{Fig: cdf} (b). Compared to the result in the sitting scenario, the performance of EyeEcho slightly decreases in all metrics.

To assess the statistical significance of this difference, we ran a repeated measure \textit{t}-test between the MAE of the sitting scenario and the walking scenario across all 12 participants and found a significant difference ($t(11)=5.51, p=0.0002<0.05$). The difference in performance is expected because walking scenarios introduced more noise caused by the motion of the user (e.g., shaking of the head) and displacement of the device. We also observed the acoustic signals were reflected differently from different background objects like walls and tables while participants were in motion. These factors collectively contribute to the slight drop in performance during user motion. Nevertheless, the results showed that our system is capable of tracking facial expressions accurately and reliably, even when the users are walking.

\begin{figure*}[t]
  \includegraphics[width=0.9\textwidth]{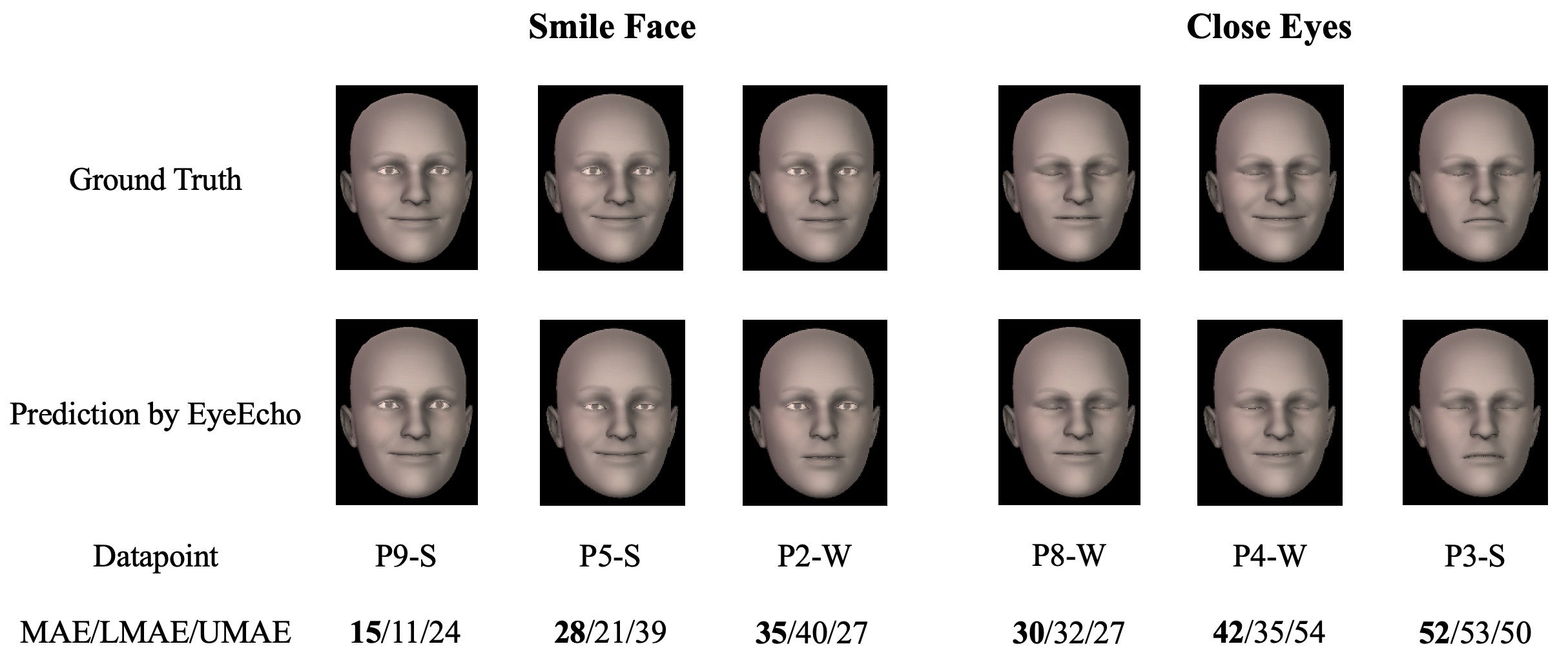}
  \caption{Visualized Results. Datapoint: P (Participant), S (Sitting), W (Walking).}
  \label{Fig: Visualized Tracking Performance}
  \Description{This figure shows the visualization of two facial expressions, Smile Face and Close Eyes. For each facial expression, we selected three frames and displayed both the ground truth expression and the predicted facial expression, along with the corresponding MAE/LMAE/UMAE. These frames were randomly selected from different participants. The prediction and the ground truth are visually highly similar to each other when LMAE is under 40 and UMAE is under 60.}
\end{figure*}

\subsubsection{Visualized Results}
\label{Subsubsec: visualized results}
To visually demonstrate the tracking performance of our system, we present the visualization results on two facial expressions, Smile Face and Close Eyes. We decided to pick 3 frames with the MAE close to 15, 25, and 35 for the smile face expression and 3 frames with the MAE close to 30, 40, and 50 for the close eyes expression. Each frame was randomly selected from the data we collected from all 12 participants. As we can see from Fig.~\ref{Fig: Visualized Tracking Performance}, a predicted frame with an LMAE under 40 and an UMAE under 60 is highly similar to the ground truth visually. Comparing this standard to our study results in Tab.~\ref{Tab: study results} with average LMAE at 20.4 and 22.7 and UMAE at 27.1 and 34.3 for the sitting and walking scenarios, our system can reliably track facial movements across different scenarios, even after remounting.

\subsection{Performance on Tracking Facial Expressions with Varying Degrees of Deformation}

Since our EyeEcho system tracks facial expressions continuously rather than classifying them, we capture the entire process of a facial expression transitioning from a neutral face to its most extreme state. In the previous subsection, we validated the overall performance of our EyeEcho system. However, it is also important to assess its performance in tracking different degrees of deformation for these facial expressions. Fig.~\ref{Fig: expressions with strength} depicts the degree of deformation of different facial parts in frames where the facial expressions are performed to the most extreme state but they usually only account for a small portion of all frames in the entire process of making facial expressions because this is a continuous process from the neutral face to the most extreme state of facial expressions and then back to the neutral face. To quantify this proportion, based on our analysis in Sec. \ref{Subsec: facial expressions}, we calculated the degree of deformation for each frame and categorized all frames in the user study into four groups based on their degrees of deformation. We present the evaluation results in terms of average MAE for each category in Tab.~\ref{Tab: study results with deformation}, along with the percentage that each category represents among all the frames. 

As shown in the table, the majority of frames in our user study have deformation levels below 150, in alignment with what has been shown in Fig.~\ref{Fig: different deformation}. The frames plotted in Fig.~\ref{Fig: expressions with strength} mostly fall into the last category in the table with the degree of deformation larger than 150, taking up less than 10\% of all the frames. Note that the maximum degree of deformation that the subtle facial movement blinking can reach is usually below 100, as shown in Fig.~\ref{Fig: expressions with strength}. Our system demonstrates satisfactory performance on tracking facial expressions for the frames with a degree of deformation smaller than 150, as Subsec.~\ref{Subsec: evaluation metrics} and Fig.~\ref{Fig: Visualized MAE} establish that the prediction and ground truth are visually highly similar when the MAE is below 40. For frames with deformation levels exceeding 150, the performance of our system is slightly lower, but it still maintains an acceptable level of accuracy, as depicted in Fig.~\ref{Fig: Visualized MAE} and Fig.~\ref{Fig: Visualized Tracking Performance}. For blinking which has more subtle degree of deformation (smaller than 100), our system also achieves promising tracking performance considering that the MAE is around 20, as compared to the visualization in Fig.~\ref{Fig: Visualized MAE}. This analysis demonstrates that our system performs well across varying degrees of deformation for different facial expressions.

\begin{table}[htbp]
\caption{Evaluation Results vs. Different Degrees of Deformation (Data Format: MAE [Percentage of Total Frames]).}
\label{Tab: study results with deformation}
\begin{tabular}{| c | c | c |} 
\hline
 Degree of Deformation & Sitting & Walking\\ [0.5ex] 
 \hline\hline
 <50 & 16.1 [8.8\%] & 17.5 [12.1\%]\\
 \hline
 50-100 & 18.8 [64.8\%] & 22.6 [62.4\%]\\
 \hline
 100-150 & 35.8 [18.0\%] & 42.8 [17.2\%]\\
 \hline
 >150 & 39.6 [8.4\%] & 43.5 [8.3\%]\\
 \hline
\end{tabular}
\end{table}

\subsection{Determining Minimum Training Data Requirement}
\begin{figure}[h]
  \includegraphics[width=0.35\textwidth]{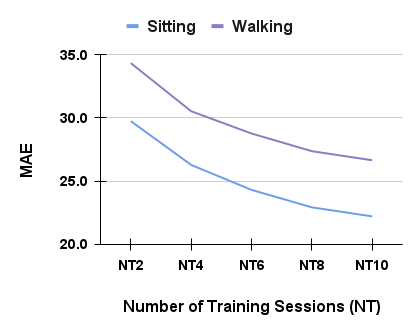}
  \caption{Impact on MAE with Different Number of Training Sessions (NT).}
  \label{Fig: nt}
  \Description{This figure shows the MAE with different number of training sessions. It shows that for both sitting and walking, more training sessions lead to better performance.}
\end{figure}

In the previous experiments, we used the data from 10 sessions (20 minutes of data) to train the model. However, 20 mins of training may not be always preferable for users in real-world deployments. Therefore, we further explore how much training data is needed before the system reaches an acceptable accuracy. We chose two sessions of data as the testing sessions and employed 2, 4, 6, 8 and 10 sessions of data respectively to train the model for each scenario. We showed the evaluation results in Fig.~\ref{Fig: nt}. As we can see in the figure, the average performance of the system improved with more training data. However, even with just 2 training sessions~(4 minutes of training data), the system achieved a MAE of 29.7 and 34.3 for the two scenarios. According to the analysis in Sec.~\ref{Subsec: evaluation metrics} and Sec.~\ref{Subsubsec: visualized results}, this result is already good enough to provide an acceptable tracking performance on facial expressions that are highly similar to the ground truth visually. In essence, two sessions of training data are likely enough to provide satisfactory tracking performance in real-world deployments for the majority of users. More training sessions could be collected for users who do not have good enough performance.

\subsection{User-Adaptive Model}
In the previous experiments, we employed a user-dependent model to predict facial expressions for each individual participant. This model was trained using the data specific to each participant. To investigate the degree of user dependency within the system, we conducted a Leave-One-Participant-Out (LOPO) experiment for each scenario. In this experiment, we utilized the data from 11 participants to train the model and then evaluated the results on the data from the remaining participant. This process was repeated for each participant, and an average result was obtained using this user-independent model.

\begin{figure}[h]
  \includegraphics[width=0.4\textwidth]{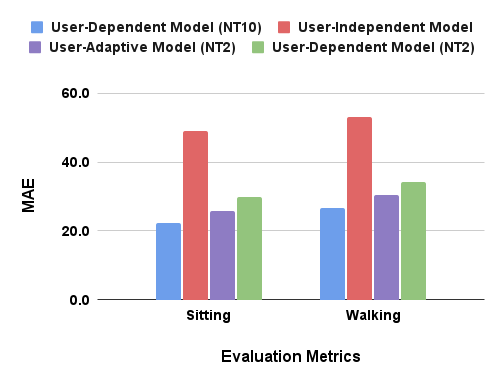}
  \caption{Impact on MAE with Different Models.}
  \label{Fig: models}
  \Description{This figure shows the MAE with different ML models. It shows that user-independent model leads to worst performance while user-adaptive model with 2 training sessions yields better performance than user-dependent model with the same amount of training data.}
\end{figure}

The results of this user-independent model are presented in Fig.~\ref{Fig: models}. Notably, in comparison to the results of the user-dependent model, the results were worse, with a MAE of 49.0 and 53.2 for the sitting and walking scenarios, respectively. However, this outcome was anticipated because our system relies on the reflection of acoustic signals on the face and head, which varies significantly among participants. Additionally, differences in how participants wore the device and executed facial expressions also contributed to this variance.

Subsequently, we explored a user-adaptive model, where we used a small portion of data collected from this participant~(2 sessions) to fine-tune this LOPO model. As shown in the figure, the results for both scenarios significantly improved compared with the user-independent model, with an MAE of 25.7 and 30.5. Besides, the results were also better than those trained using a user-dependent model with the same amount of training data (2 sessions), with an MAE of 29.7 and 34.3. This result showed the potential for further performance enhancement. If the model can be trained with a much larger data set from more participants in the future, the performance of the model can be further improved with minimal training data from a new user. 

\subsection{Transfer Learning using Data from Sitting to Walking Scenario}
All the experiments and analysis above separated the sitting and walking scenarios and reported the performance independently. This indicates a new user is required to provide data for both sitting and walking scenarios, which may not offer the optimal user experience. In this experiment, we explored using the data collected from the sitting scenario to train a model, which is transferred and evaluated on the testing data collected in the walking scenario. For this experiment, we compared three models. Among the three types of models, the testing data was the same two sessions collected in the walking scenario. The results were averaged across all 12 participants.

\begin{figure*}[t]
  \includegraphics[width=0.6\textwidth]{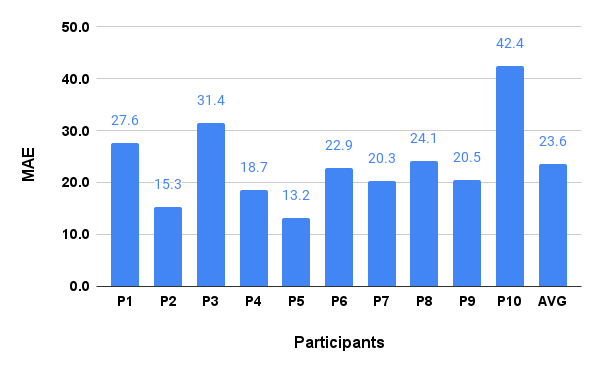}
  \caption{MAE for all Participants in the Study with Operating Frequency Range 20-24 kHz.}
  \label{Fig: MAE for frequency study}
  \Description{This figure demonstrates the average MAE for all 10 participants in the study with the operating frequency range set as 20-24 kHz. They are 27.6, 15.3, 31.4, 18.7, 13.2, 22.9, 20.3, 24.1, 20.5, 42.4 for P1-P10. The average is 23.6.}
\end{figure*}

The first model is User-dependent Model (NT10), where 10 sessions from the walking scenario were used to train the model with no transfer learning applied. The average MAE for this model is 26.9. In the second model (Transfer learning Model (NT2)), the model was trained with 10 sessions of data from the sitting scenario and fine-tuned using 2 sessions of data from the walking scenario. This model yields an average MAE of 29.0. In the third model, only 2 sessions from the walking scenario were used to train the model. The average MAE in this case is 34.3. The results show that transfer learning improves the MAE from 34.3 to 29.0, using the same size of training sessions from the walking scenario. It shows that EyeEcho has the potential to be adapted to the new walking scenario with minimal training data needed. We plan to explore how to use advanced transfer learning to further reduce the training data for new scenarios in the future.

\section{Evaluation of EyeEcho with Different Settings}\label{Sec: evaluation with different settings}
In this section, we conducted a new study in the lab to evaluate EyeEcho's performance with different operating frequency ranges and under different noisy environments, as well as the usability of EyeEcho.

\subsection{Impact of Operating Frequency Range}\label{Subsec: frequency impact}
The EyeEcho system was designed to operate within the frequency range of $16-20 kHz$, with the goal of easy adoption on most commodity speakers and microphones since the acoustic sensors in most commodity devices can sample up to 48KHz. In post-study surveys, participants did not report any issues with hearing the acoustic signals. While it is possible that some users may be able to hear it, EyeEcho can readily adapt to higher inaudible frequencies with minimal impact on tracking performance. In order to validate this assumption, we shifted the operating frequency range of the EyeEcho system to $20-24 kHz$ and conducted a new in-lab user study of the same procedure described in Sec.~\ref{Subsec: study design} with 10 participants (3 females and 7 males, 22 years old on average). With the 12 sessions of data we collected from each participant, we also ran a 6-fold cross-validation by using 10 sessions to train the model and 2 sessions to evaluate the performance. The average MAE of each participant and all 10 participants on average are demonstrated in Fig.~\ref{Fig: MAE for frequency study}.

As shown in Fig.~\ref{Fig: MAE for frequency study}, the average MAE of all 10 participants in this study is 23.6, which is comparable to the MAE of 22.9 in the first study in Sec.~\ref{Sec: user study} with the operating frequency range set at $16-20 kHz$. This validates that EyeEcho can be adapted to a higher inaudible frequency range ($20-24 kHz$) with little impact on the system performance. Furthermore, we specifically asked each participant "Can you hear the sound emitted from our system? Yes / No" in the questionnaires collected at the end of this study and all 10 participants answered 'No' to this question. Therefore, we believe that EyeEcho can operate at a frequency range that has minimal impact on users' daily activities with a satisfactory tracking performance.

\begin{figure*}[t]
  \includegraphics[width=\textwidth]{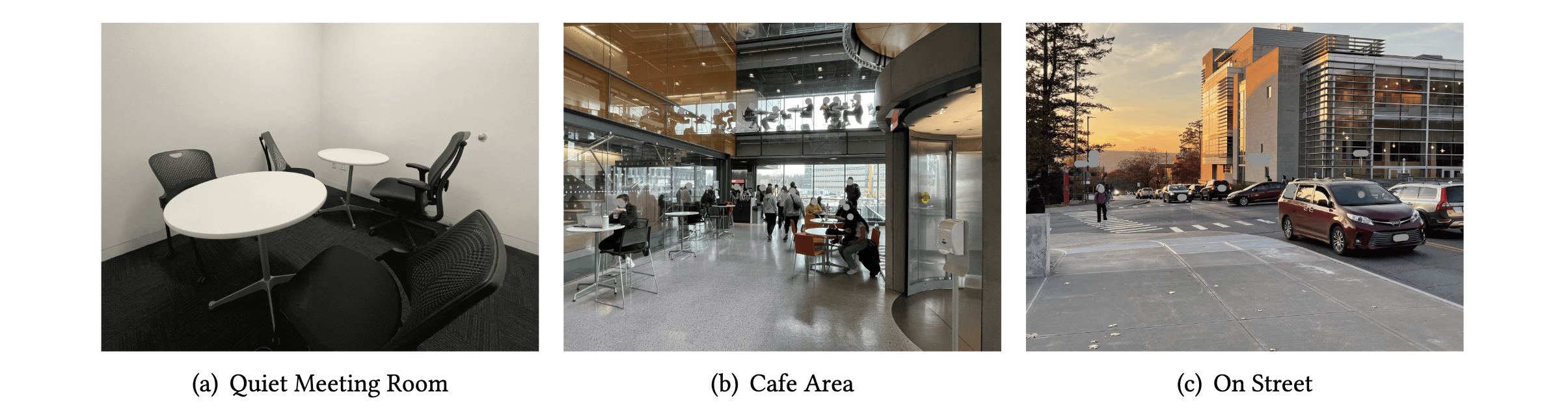}
    \caption{Three Different Noisy Environments.}
    \label{Fig: noisy environments}
    \Description{This figure shows three noisy environments. (a) is a quiet meeting room with chairs and tables, (b) is a cafe area with many people working and talking and (c) is a crossroad on the street with vehicles and pedestrians passing by.}
\end{figure*}

\begin{table*}[htbp]
\caption{Evaluation Results in MAE for Different Noisy Environments.}
\label{Tab: noisy results}
\begin{tabular}{| c | c | c | c | c | c | c | c | c | c | c | c |} 
\hline
 Environments & P1 & P2 & P3 & P4 & P5 & P6 & P7 & P8 & P9 & P10 & \textbf{AVG}\\ [0.5ex] 
 \hline\hline
 Quiet Meeting Room (40.6 dB) & 31.3 & 16.7 & 26.0 & 20.7 & 13.4 & 24.9 & 20.1 & 29.4 & 20.0 & 39.1 & \textbf{24.2}\\
 \hline
 Play Music (64.5 dB) & 35.1 & 27.6 & 32.0 & 17.2 & 15.2 & 29.6 & 29.8 & 23.3 & 26.6 & 36.4 & \textbf{27.3}\\
 \hline 
 In Cafe (56.9 dB) & 38.7 & 22.1 & 38.2 & 24.5 & 17.6 & 25.9 & 27.0 & 23.4 & 28.0 & 41.6 & \textbf{28.7}\\
 \hline
 On Street (69.3 dB) & 47.7 & 41.9 & 62.1 & 51.7 & 42.9 & 46.8 & 44.6 & 39.8 & 33.8 & 54.2 & \textbf{46.6}\\
 \hline
\end{tabular}
\end{table*}

\subsection{Impact of Environmental Noises}
\label{Subsec: noise impact}
Since EyeEcho uses acoustic signals as the sensing method, there is a chance that everyday environmental noise could have a negative impact on the tracking performance.
To investigate how different environmental noises can affect performance, we further extended the new study in Sec.~\ref{Subsec: frequency impact}. The study was originally conducted in a quiet meeting room with the background noise of the air conditioner, as shown in Fig.~\ref{Fig: noisy environments} (a). To evaluate the EyeEcho system under different noisy environments, after the first part study was completed in the quiet room, we then asked the participants to move to different environments to collect testing data with the existence of various types of noises: (1) \textit{Music Noise}: in the experiment room with random music played (Fig.~\ref{Fig: noisy environments} (a)); (2) \textit{Cafe Noise}: in a cafe with cafe staff and customers talking (Fig.~\ref{Fig: noisy environments} (b)); (3) \textit{Street Noise}: on the street near a crossroad with vehicles and pedestrians passing by (Fig.~\ref{Fig: noisy environments} (c)). In every one of the three noisy environments above, each participant performed facial expressions for 2 sessions (4 minutes) as testing data.

Then, we used the first 10 of 12 sessions of data collected in the quiet room to train a model, which was tested using the remaining 2 sessions of data in the quiet room and 2 sessions of testing data collected in different noisy environments. Please note that no data collected in the noisy environments was used for training. The evaluation results are displayed in Tab.~\ref{Tab: noisy results}. We also measured the noise level in each noisy environment for each participant and showed the average measurement in Tab.~\ref{Tab: noisy results}.

The evaluation results in Tab.~\ref{Tab: noisy results} demonstrate that the average MAE for 10 participants remains consistent at 27.3 and 28.7 with the presence of music noise and cafe noise compared with the MAE of 24.2 in the quiet environment. There is a small performance variance for some participants among these three environments because testing data is not large-scale but overall the system is resistant to these two noises considering that the prediction is visually similar to the ground truth when the MAE is below 40 as discussed in Sec.~\ref{Subsec: evaluation metrics}. However, the system performance dropped significantly when the user study was conducted on the street. We further explore the possible causes below.

\begin{figure*}[t]
  \includegraphics[width=\textwidth]{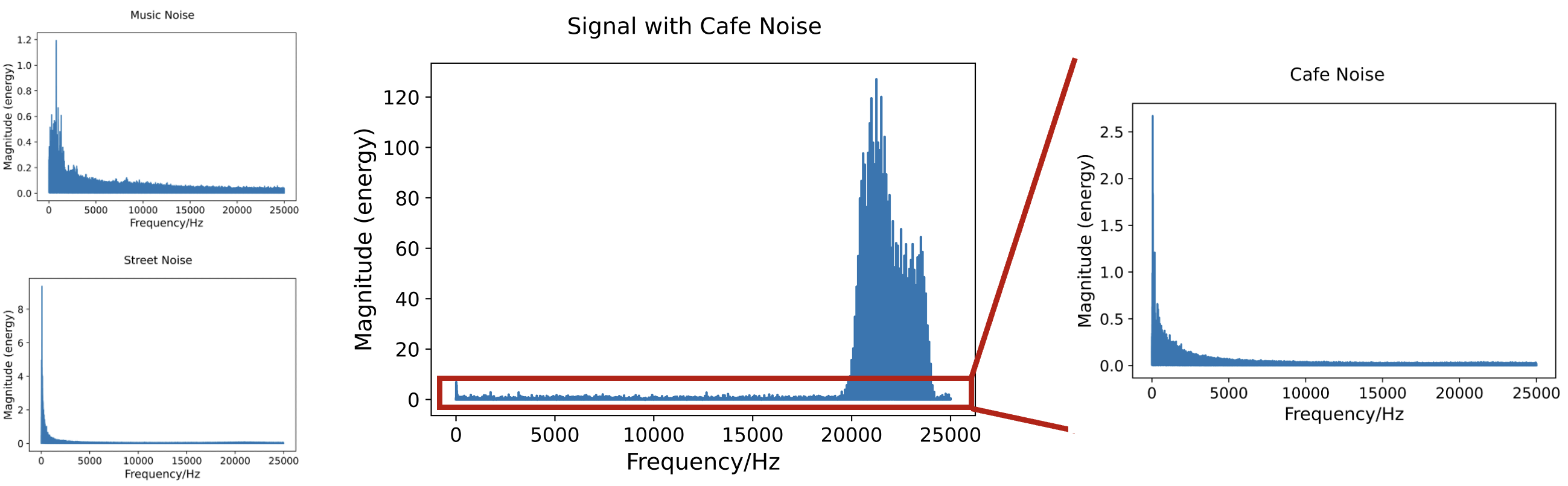}
  \caption{Signal with Noises in Frequency Domain and Zoom-in Plots of Noises.}
  \label{Fig: signal with noises}
  \Description{This figure shows the signal in our system together with the cafe noise in the frequency domain. Besides, it also shows three noises, which are cafe noise, music noise and street noise, in the frequency domain. The noises are mainly in the audible frequency range (under 5 kHz) and the strength of them are much smaller compared to the signal.}
\end{figure*}

\begin{figure*}[t]
  \includegraphics[width=\textwidth]{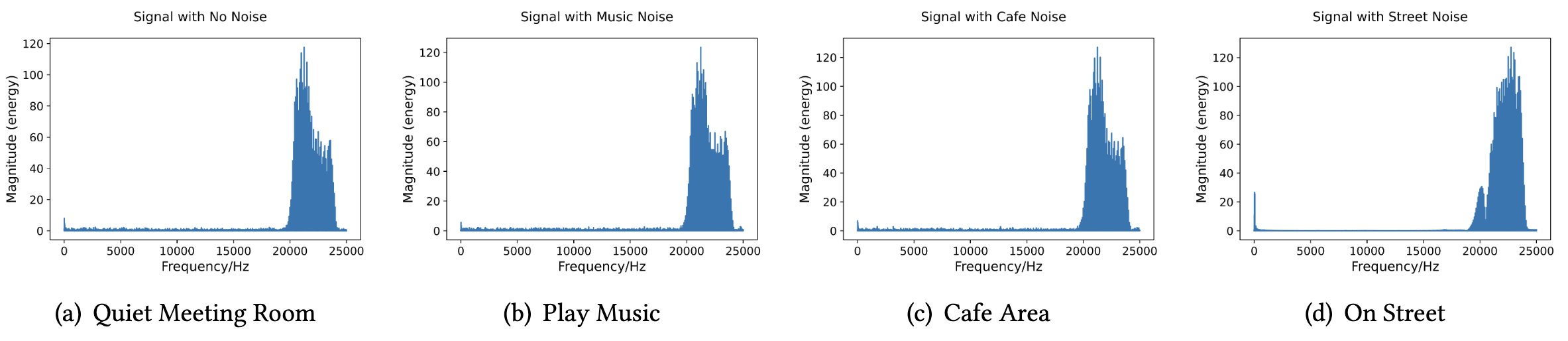}
    \caption{Signal in Frequency Domain under Four Noisy Environments. All figures are plotted using the data collected with P5.}
    \label{Fig: signal in different environments}
    \Description{This figure displays the signal with no noise, music noise, cafe noise and street noise in the frequency domain from P5's data. The first three signals look very similar while the shape of the signal with street noise is very different.}
\end{figure*}

We first plotted both the signal with noises and the pure noises in the frequency domain in Fig.~\ref{Fig: signal with noises}. As we can see in the figure, all three noises are mostly within the audible frequency ranges and will be filtered out by the band-pass filter in our system. Besides, the strength of the noises is much smaller than the signal because the sources of these noises are relatively far away from the microphones in our system. This helps explain why the music noise and the cafe noise have little impact on the system performance. In theory, street noise should also have a limited impact on the system's performance. However, the evaluation results suggest otherwise. Hence, we further analyzed the received acoustic signal in the frequency domain in different noisy environments. We used P5's data as an example for illustration and plotted the signals under different noisy environments since the signal patterns are similar among all participants. As shown in Fig.~\ref{Fig: signal in different environments}, the signals are very similar to each other in the frequency domain under the first three environments while the signal looks quite different when collected on the street. Fig.~\ref{Fig: signal with noises} already shows that the noises have a limited impact on the signal so we believe this difference is mainly caused by the temperature difference between indoor and outdoor environments. According to prior research~\cite{hayashida2020estimation,micenvironment}, the frequency response of both speakers and microphones can be largely impacted by the temperature of the environment where they operate. The datasheets of the speakers~\cite{sr6438} and the microphones~\cite{ics43434} used in our system suggest that the speakers can be more vulnerable to the temperature change. Our study was conducted in a cold region where the outdoor temperature varied from $-5\degree C$ to $5\degree C$ when the testing data on the street was collected for 10 participants while the indoor room temperature remained between $20\degree C$ to $25\degree C$.

\begin{figure*}[t]
  \includegraphics[width=\textwidth]{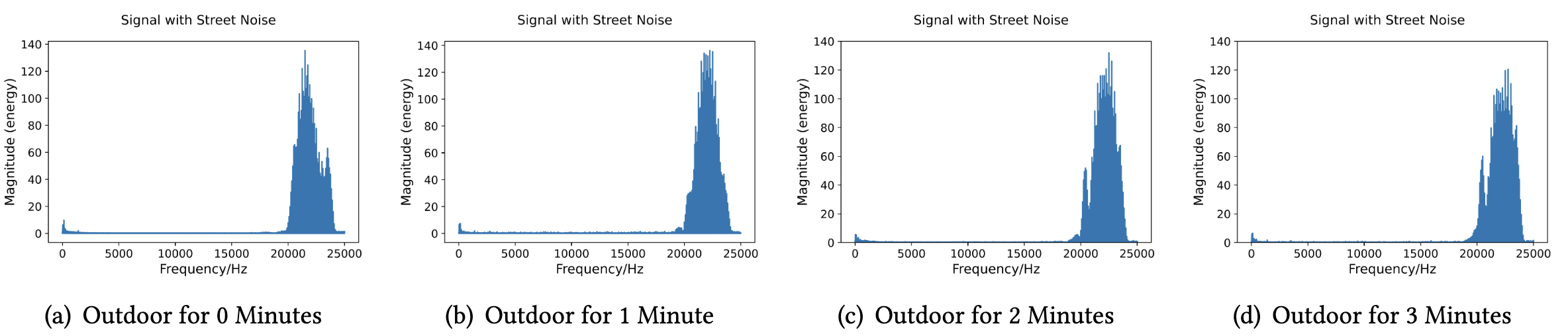}
    \caption{Signal in Frequency Domain at Different Time after EyeEcho being Moved from Indoor ($22\degree C$) to Outdoor ($1\degree C$).}
    \label{Fig: signal in different temperature}
    \Description{The figure shows the signal change after the device was moved from indoor to outdoor. 1 minute after the device was moved outdoor, the signal shape changed. 2 minutes and 3 minutes after, the signal shape continued to change and became stable.}
\end{figure*}

To further verify this hypothesis, 
we did one experiment: we moved the EyeEcho system from the indoor environment ($22\degree C$) to the outdoor environment ($1\degree C$) and recorded the received signal with our system at different times after the system was brought outdoors. The change of the received signal in the frequency domain is presented in Fig.~\ref{Fig: signal in different temperature}. As the figure shows, the received signal in the EyeEcho system visually changed after the temperature decreased. Since only the on-street testing data was collected outdoors, this change only affected the system performance for the testing data collected on the street, because the training data and testing data are significantly different in this condition.

In summary, our EyeEcho system is robust to different kinds of daily noises because it operates in the ultrasonic band. However, our system may need further calibration in the areas where the temperature is significantly low. This can be achieved by choosing sensors whose frequency response is more resistant to temperature changes and collecting more training data under different temperatures. We will explore this in the future.

\subsection{Usability of EyeEcho}\label{Subsec: questionnaire}

To explore the usability of the EyeEcho system, participants were requested to finish a questionnaire at the end of the study in Sec.~\ref{Subsec: frequency impact}. The participants first rated their overall experience with EyeEcho by answering two questions: (1) \textit{"How comfortable is this wearable device to wear around the face? (0 most uncomfortable, 5 most comfortable)"}; (2) \textit{"How acceptable do you find the weight of our wearable device? (0 most unacceptable, 5 most acceptable)"}. On average, 10 participants gave scores of 4.2 and 4.8 to the two questions above. All 10 participants agreed with the statement "The pair of glasses is easy to use." except that P9 reported that "glasses would fall in a squeeze gesture". As for the question "Compared with normal glasses, what do you think of this device?", 9 participants thought that EyeEcho is generally very similar to a pair of normal glasses while P1 thought it is a little bit more hard to be put on than normal glasses because the legs of the glasses cannot be bent. Meanwhile, P3 and P10 suggested that EyeEcho could have selected prescription lenses for different users in future while P4 believed that it will be easier to wear EyeEcho if all sensors are completely embedded into the legs of the glasses. We believe that all these suggestions are valuable and we will take them into consideration in future improvement of the prototype design.

\begin{figure*}[t]
    \centering
    \subfloat[Living Room]{
        \includegraphics[width=5cm]{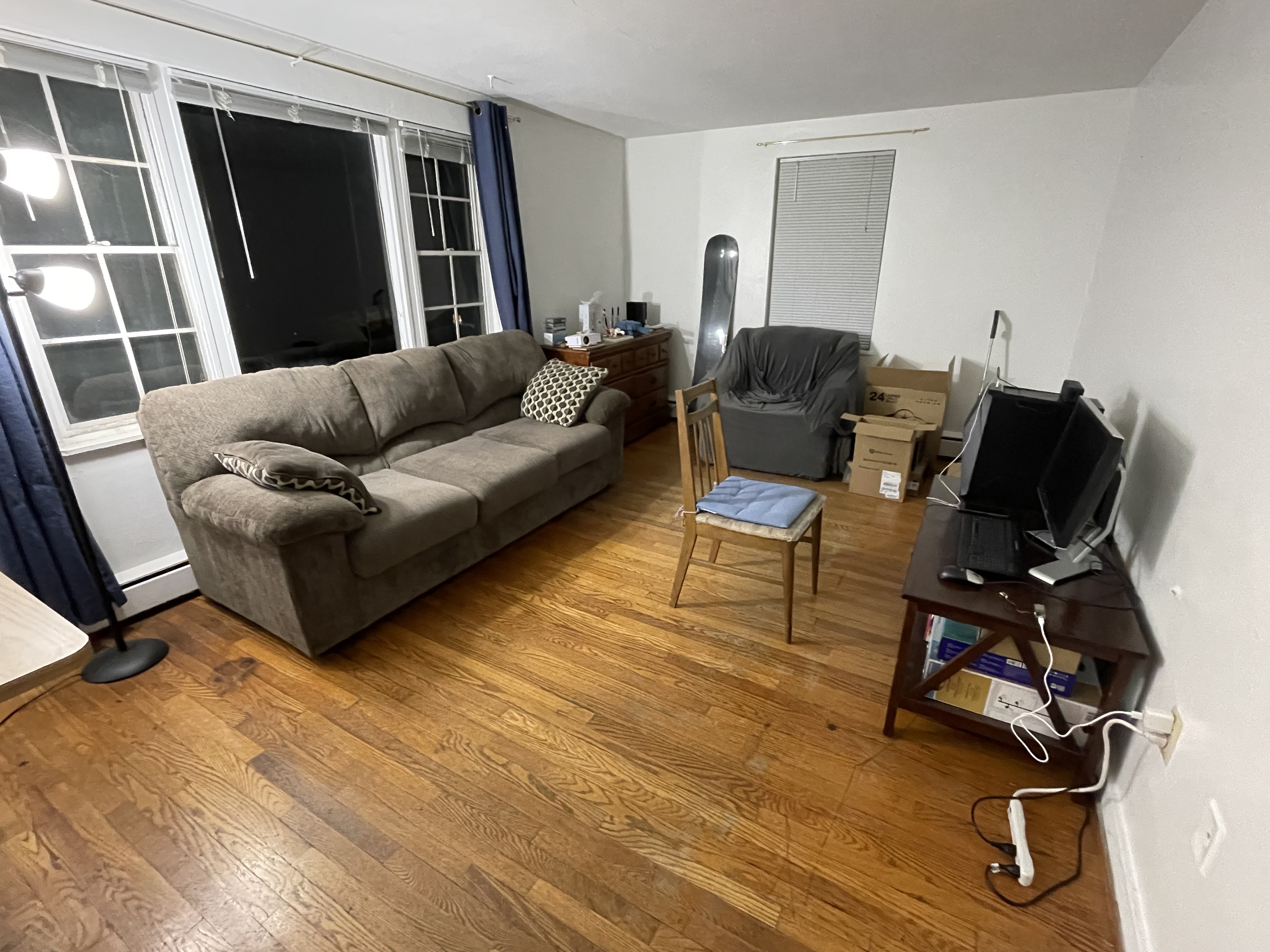}
    }
    \subfloat[Bedroom]{
        \includegraphics[width=5cm]{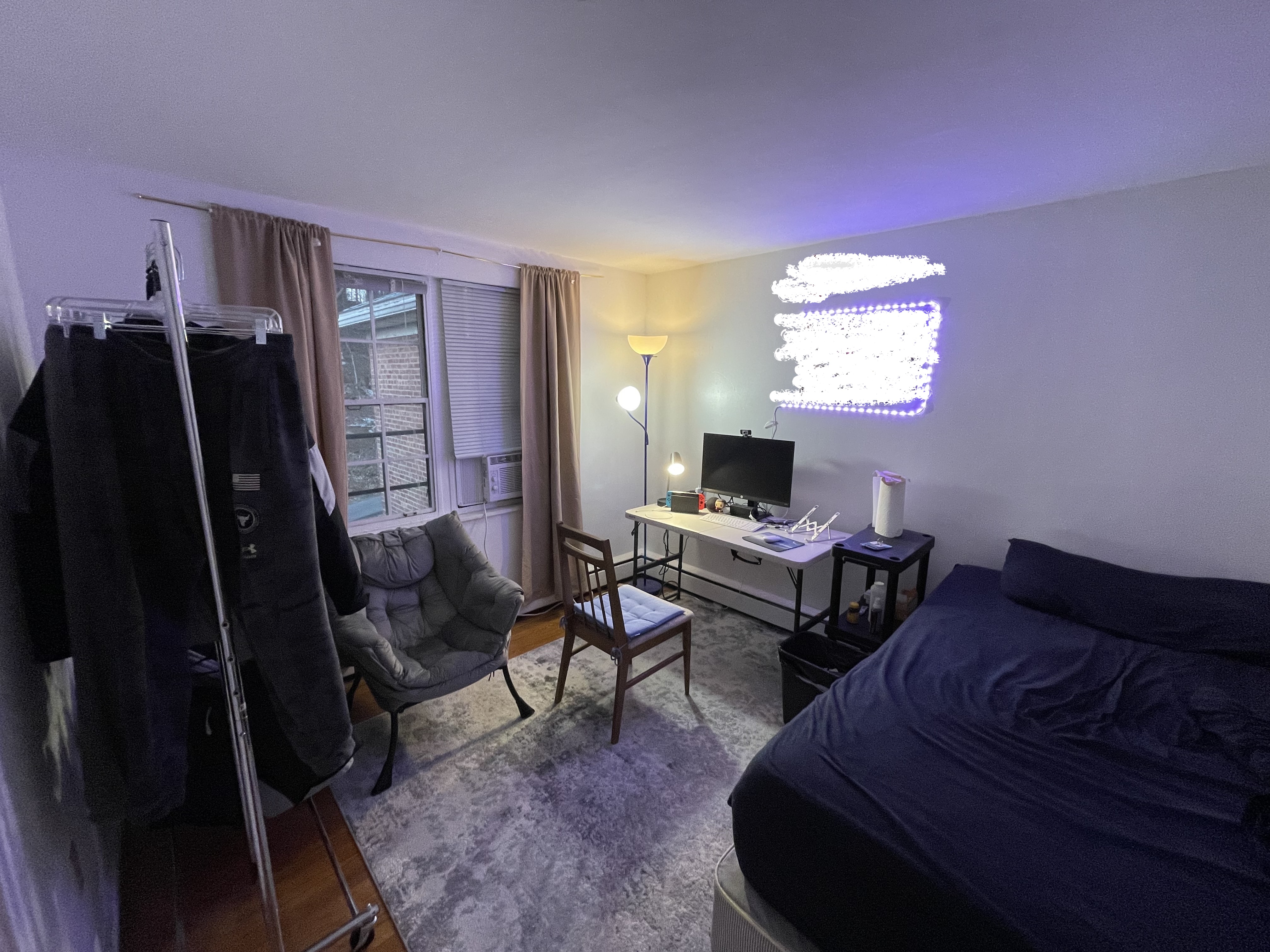}
    }
    \subfloat[Kitchen]{
        \includegraphics[width=5cm]{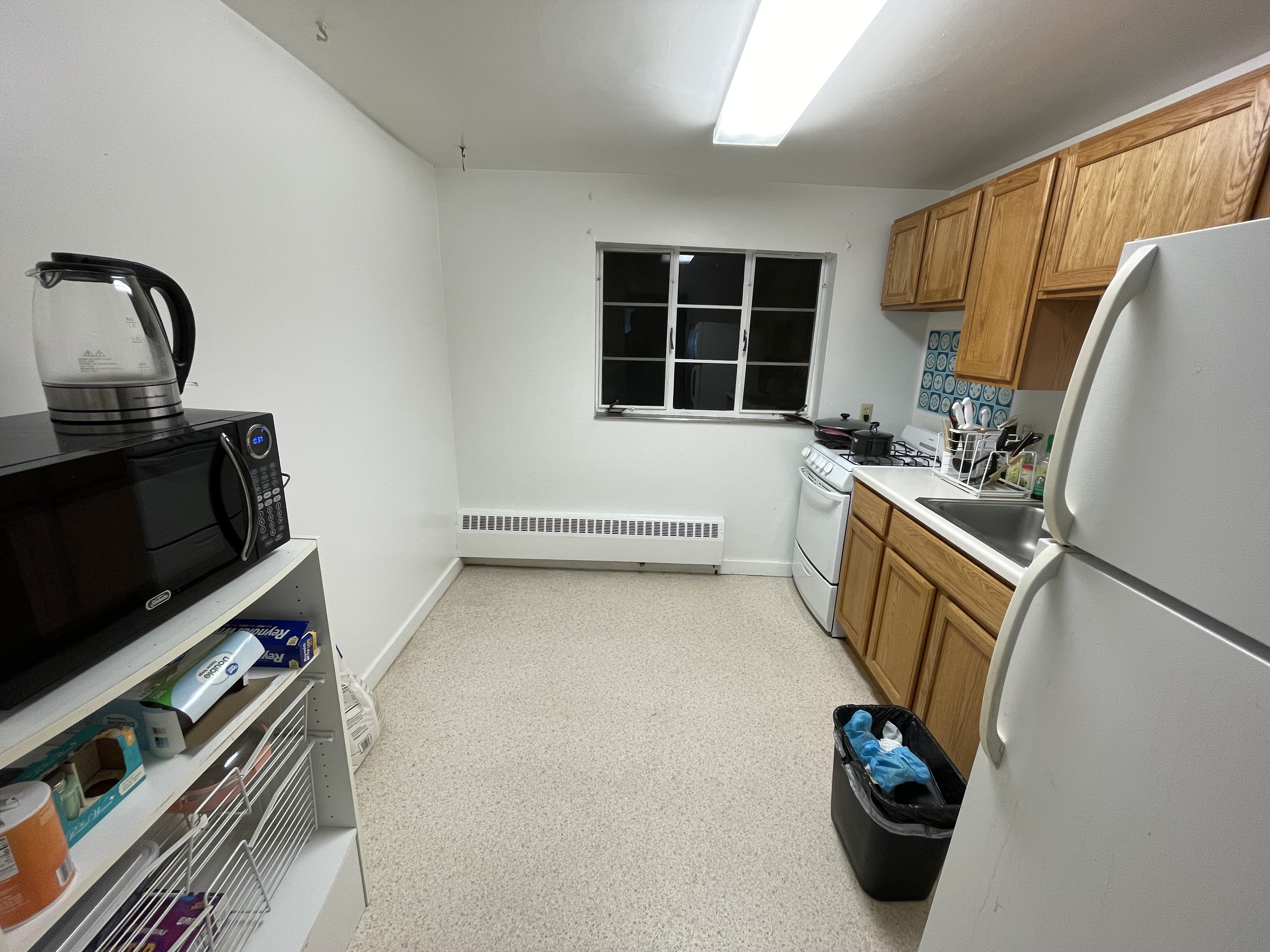}
    }
    \caption{Apartment used for the Semi-in-the-wild Study.}
    \label{Fig: apartment}
    \Description{This figure shows the furniture layout of three rooms, living room, bedroom and kitchen in a one-bedroom apartment.}
\end{figure*}

At last, we asked participants that "If there is a product like this to track your facial expressions in future, do you want to use it?". 7 participants answered 'Yes' to this question while 2 participants answered 'Maybe". P8's choice is dependent on the performance of the system and he "may consider this device if it can perform nearly as well as existing technology".
\section{Evaluation of EyeEcho in a Semi-in-the-wild Study}\label{Subsec: wild study}
\subsection{Study Design}

Evaluating a facial expression tracking system in real-world environments presents significant challenges, primarily due to the absence of a suitable method for acquiring ground truth data that users can comfortably wear during their daily activities. For instance, prior research, as well as our own study for both sitting and walking scenarios in the lab, have relied on placing a camera in front of the users' face to capture ground truth data of facial expressions. However, it becomes nearly impractical to expect participants to wear a ground truth acquisition device in a completely uncontrolled environment, where they have the freedom to go to any location (especially outdoor locations) and engage in any activity (especially activities such as driving) without the presence of researchers.

The lack of reliable and minimally-obtrusive wearable systems that can track users' facial expressions continuously has also been a significant motivator behind the development of EyeEcho. In our initial study, we successfully showcased the promising performance of EyeEcho in a controlled lab setting, where we simulated various real-world scenarios, including factors like motion and noise. Based on the exciting results, we also would like to demonstrate that our proposed system can perform effectively in a more naturalistic setting since our core sensing principle, which involves tracking inaudible acoustic reflections on the face, is less susceptible to environmental influences.

To validate our hypothesis, we designed and conducted a second study, referred to as a semi-in-the-wild user study in which participants engaged in various daily activities within a more naturalistic setting—a one-bedroom apartment. In this study, we aim to evaluate EyeEcho in an environment as natural as the real-world setting, while ensuring that the ground-truth acquisition system works well. To the best of our knowledge, this study marks the first attempt in the field to evaluate a non-camera-based wearable facial expression tracking system in a more naturalistic real-world setting instead of controlled lab settings.

The primary goals of this study are as follows:
\begin{itemize}
\item Evaluate the performance of EyeEcho in multiple rooms within a home environment, where furniture and layouts vary;
\item Assess how well EyeEcho can track natural facial expressions that occur during various daily activities.
\end{itemize}

\subsection{Study Environment and Activities}
The study was conducted in a one-bedroom apartment of a researcher off campus, with three rooms: the living room, bedroom, and kitchen, as depicted in Fig.~\ref{Fig: apartment}. The participants were instructed to perform different activities in a random order while wearing the experimental devices within these rooms, as follows:

\begin{itemize}
    \item Living room: Watching videos on a computer, reading, describing things, having conversation with the researcher while walking around;
    \item Bedroom: Watching videos on a computer, reading, describing things, having conversation with the researcher while making the bed;
    \item Kitchen: Watching videos on a computer, describing things, having conversation with the researcher while washing the dishes and using the microwave oven.
\end{itemize}

These activities were intentionally designed to elicit a range of natural facial expressions and movements that happen in everyday life. For example, the videos that participants watched included online videos categorized to specifically evoke different emotional experiences\footnote{\url{https://www.alancowen.com/video}} and pre-selected YouTube videos consisting of various funny/scary scenes that happened in a movie or in real life. Watching these videos led to spontaneous and varied facial expressions. While reading, describing things and having conversations, participants had facial movements frequently. According to the standard defined in Sec.~\ref{Subsec: facial expressions}, among all the frames collected in this user study, there are 84.4\% and 24.6\% frames in which participants deformed their face with a degree of deformation over 50 and 100, respectively. This is comparable to the degrees of deformation that the participants performed in the in-lab study.

\begin{figure*}[t]
    \centering
    \subfloat[MAE vs. Participants (without fine-tuning)]{
        \includegraphics[width=9.5cm]{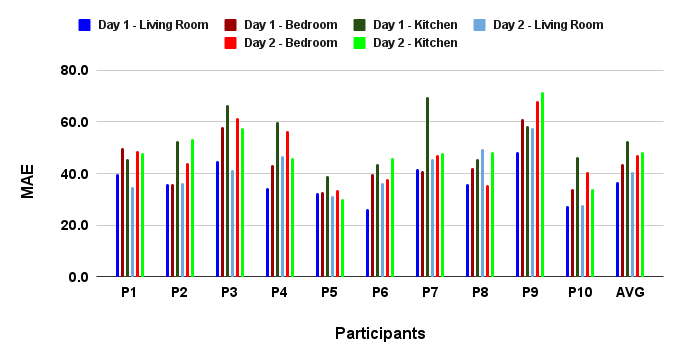}
    }
    \subfloat[MAE vs. Experiment Rooms (without and with fine-tuning)]{
        \includegraphics[width=6.2cm]{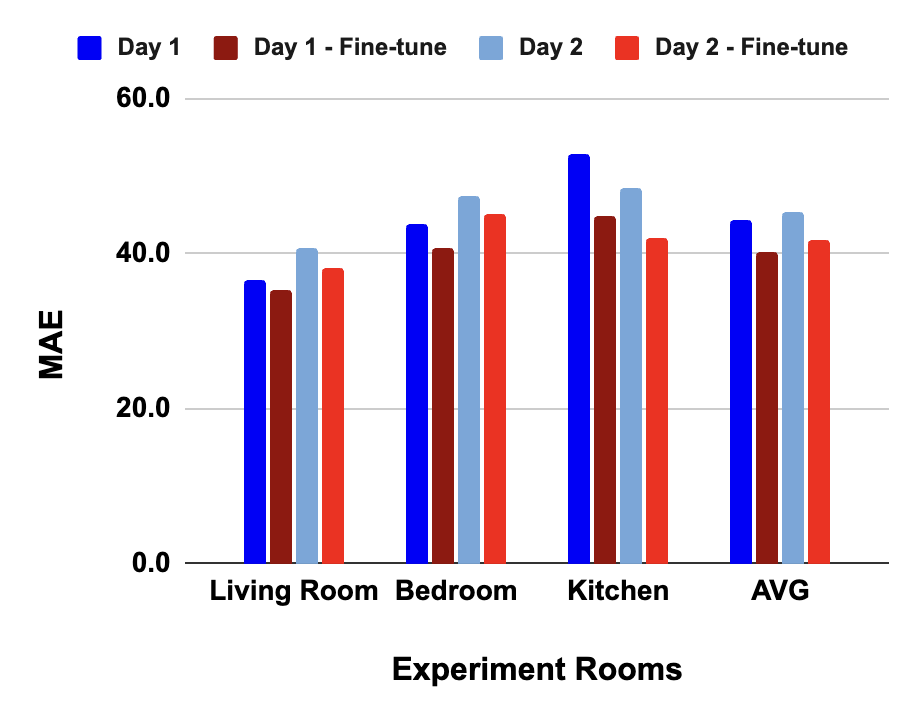}
    }
    \caption{Performance of EyeEcho for the Semi-in-the-wild Study.}
    \label{Fig: wild results}
    \Description{This figure shows performance of our system for the semi-in-the-wild study. Subfigure (a) shows MAE without fine-tuning for all 10 participants and average in the living room, bedroom and kitchen. Living room has the best performance and the kitchen has the worst performance. Generally, Day 1 has better performance than Day 2. Subfigure (b) shows MAE with and without fine-tuning. Fine-tuning improves performance for all three rooms and both days, especially for kitchen.}
\end{figure*}

\subsection{Study Procedure}
For this semi-in-the-wild study, we recruited 10 participants (8 female, 2 male) with an average age of 23 years. Each participant received USD \$20 compensation for each study day. The study was conducted in the one-bedroom apartment as detailed above.

Throughout the study, the participants wore the glasses embedded with the EyeEcho system and a chest mount, as used in the walking study, to facilitate ground truth capture via an iPhone placed in front of them. The chest mount was tested for comfort and usability before the study. Each participant completed the study over two days with a gap less than one week, engaging in various activities as described earlier. On the first day, participants conducted a 10-minute training session in the living room, followed by 10-minute testing sessions in all three rooms. The order of the rooms was randomized. On the second day, participants completed 10-minute testing sessions in all three rooms, with no additional training data collected. In total, each participant contributed approximately 70 minutes of data (10 minutes training on Day One, 60 minutes testing on both days). The entire study duration for each participant did not exceed 2.5 hours.

In the study, a researcher remained outside the experiment room to provide instructions and engage in conversations with participants via smartphone, simulating real-world scenarios where users may have video conferences while multitasking or moving around without a camera continuously in front of them.

\subsection{Study Results}

\subsubsection{Evaluation Protocol}
To predict the facial expressions performed by participants in the semi-in-the-wild study, we utilized the same deep learning model described in Sec.~\ref{Subsec: learning algorithm}. Initially, we trained a large base model using all the data collected during the in-lab study, as outlined in Sec.~\ref{Sec: user study}. This included data from both sitting and walking scenarios. Please note that there was no overlap between the two groups of participants in the two studies. Incorporating more data resulted in improved performance compared to solely using data collected in the second user study (semi-in-the-wild) as training data, based on our preliminary experiments in pilot studies. Subsequently, we conducted further training on the large base model using the 10-minute training data collected in the living room on the first day for each participant. 

\subsubsection{Data Augmentation for Enhanced Model Robustness}

During the training process, we implemented two data augmentation methods to enhance the robustness of our system. 1) Firstly, we applied random vertical shifts to the input differential echo profiles to mitigate the impact of device remounting; 2) Secondly, we introduced random walking patterns, collected by a researcher, into the training data to augment the model's ability to make predictions while participants were walking. These data augmentation techniques were employed to improve the model's performance and ensure its adaptability to varying conditions and scenarios.

\subsubsection{Results}

The evaluation results across different participants are shown in Fig.~\ref{Fig: wild results} (a). On average, the MAE across 10 participants are 44.4 and 45.5 for Day One and Day Two. Separately, the MAE are 36.7, 43.8 and 52.7 for the living room, the bedroom and the kitchen on Day One and are 40.7, 47.4 and 48.3 for these three rooms on Day Two. This performance was achieved when we only collected 10-minute training data in the living room on Day One for each participant. The living room has the best performance because the training data was only collected in it. The kitchen has a relatively worse performance because the activities performed in it had more differences from those performed in the living room. Please note that P9 has the worst performance among all participants because this participant wore a Hijab (a head covering) during the study which we believe partially blocked the transmission and reflection of the signals. We ran a repeated measures \textit{t}-test between the average MAE of three rooms on two different days across 10 participants and did not find a statistically significant difference ($t(9)=0.79, p=0.45>0.05$). This proves that the performance of our system maintains solid across different days and users do not have to collect new training data on different days.

\begin{figure*}[t]
  \includegraphics[width=0.8\textwidth]{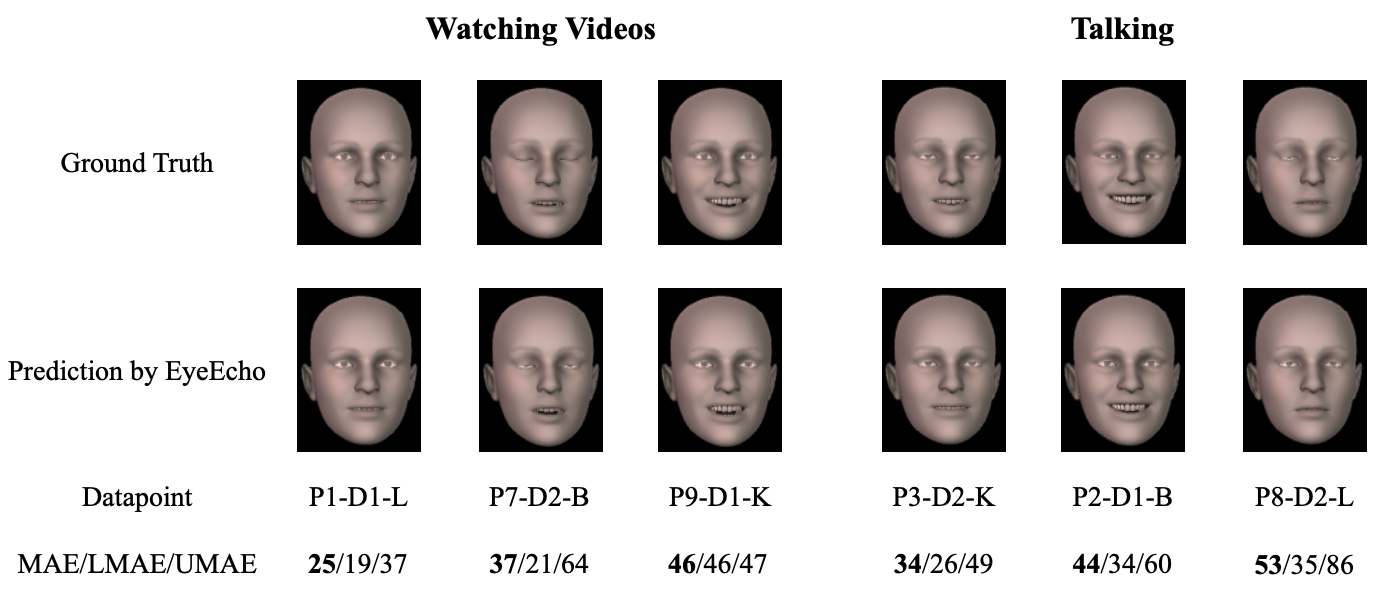}
  \caption{Visualized Results for Semi-in-the-wild Study. Datapoint: P (Participant), D (Day), L (Living Room), B (Bedroom), K (Kitchen).}
  \label{Fig: Visualized Wild Performance}
  \Description{This figure shows the visualization of results in the semi-in-the-wild study. We displayed the performance for two activities, watching videos and talking, and selected three frames for each activity. The frames were randomly selected from all participants. The frames are marked with corresponding MAE/LMAE/UMAE. This figure shows that the prediction and the ground truth are visually similar to each other when LMAE is under 40 and UMAE is under 60.}
\end{figure*}

\subsubsection{Fine-tuned results}
To further boost the performance of the system, we used 30-second data at the beginning of each 10-minute session to fine-tune the trained model for this session. The comparison between results with and without fine-tuning is demonstrated in Fig.~\ref{Fig: wild results} (b). As shown in the figure, fine-tuning improved the overall MAE from 44.4 to 40.3 for Day One and from 45.5 to 41.7 for Day Two. Specifically: (1) for the living room, MAE was improved from 36.7 to 35.3 on Day One and from 40.7 to 38.2 on Day Two; (2) for the bedroom, it was improved from 43.8 to 40.7 on Day One and from 47.4 to 45.1 on Day Two; (3) for the kitchen, it was improved from 52.7 to 44.7 on Day One and from 48.3 to 41.9 on Day Two. The fine-tuning mainly improves the performance of our system in a new room where no training data was collected before and it can be done only once in this room. Although fine-tuning with 30-second data improves the performance of our system, it might impact users' experience when they use our system in the real world. However, we believe that the results without fine-tuning (44.4 for Day One and 45.5 for Day Two on average) are also acceptable even though they are not as good as the in-lab results because the prediction is still visually similar to the ground truth when MAE is around 40 according to the analysis in Sec.~\ref{Subsec: evaluation metrics} and Sec.~\ref{Subsubsec: visualized results}. If we can collect more training data in various scenarios from more participants in the future, the performance of our system can be further improved even without the fine-tuning process.

\subsubsection{Visualized Results}

To help readers better understand the performance of our system, we visually illustrate its ability to track facial expressions in the semi-in-the-wild study, as we did in Sec.~\ref{Subsubsec: visualized results}.  We selected frames with different MAE from the data we collected in this study to show the visualized output results of facial expressions. We selected 3 frames from each of the two typical activities during which participants had frequent facial expressions, watching videos and talking. The results are shown in Fig.~\ref{Fig: Visualized Wild Performance}.

As we can see in Fig.~\ref{Fig: Visualized Wild Performance}, a prediction with an MAE of around 40 is visually similar to the ground truth. According to Fig.~\ref{Fig: wild results} (b), our EyeEcho system reaches an average MAE of 44.4 and 45.5 for Day One and Day Two without fine-tuning. This validates the performance of our system in this semi-in-the-wild study, where users conducted activities in different rooms and used the device on different days. Furthermore, in this study, participants played the sound while they were watching videos and created loud noise while they were using the microwave oven and washing the dishes. This confirms that our system is not easily impacted by common daily noises.
\section{Discussion}
\label{Sec: Discussion}

\subsection{Power Consumption Analysis}
We used a current ranger and a multimeter to measure the current and voltage of our EyeEcho system while all components were in operation. The measurements showed that the current flowing through our system was $41.1 mA$ at the voltage of $4.07V$. Therefore, the power consumption of EyeEcho is around $167 mW$. This power consumption should allow EyeEcho to work on current smart glasses or AR glasses for a reasonable period of time.

Smart glasses often come with limited battery size due to their compact device size. For instance, Amazon Echo Frame with 4 speakers can last about 2 hours with audio on~\cite{echoframesbattery}. The Ray-ban Stories Smart Glasses have a battery capacity of $167 mAh$~\cite{raybanglasses}. If EyeEcho is deployed on it and used alone, the glasses can last 4 hours in theory. On the other hand, AR glasses often come with a larger size and battery life. For instance, the battery capacity of Google Glass, Espon Moverio, and Microsoft HoloLens are $570 mAh$~\cite{googleglass}, $3400 mAh$~\cite{Moverio} and $16500 mAh$~\cite{hololensspec}, guaranteeing around 14, 83, and 402 hours of battery life in theory, if EyeEcho is used alone.

\begin{figure}[h]
  \includegraphics[width=0.3 \textwidth]{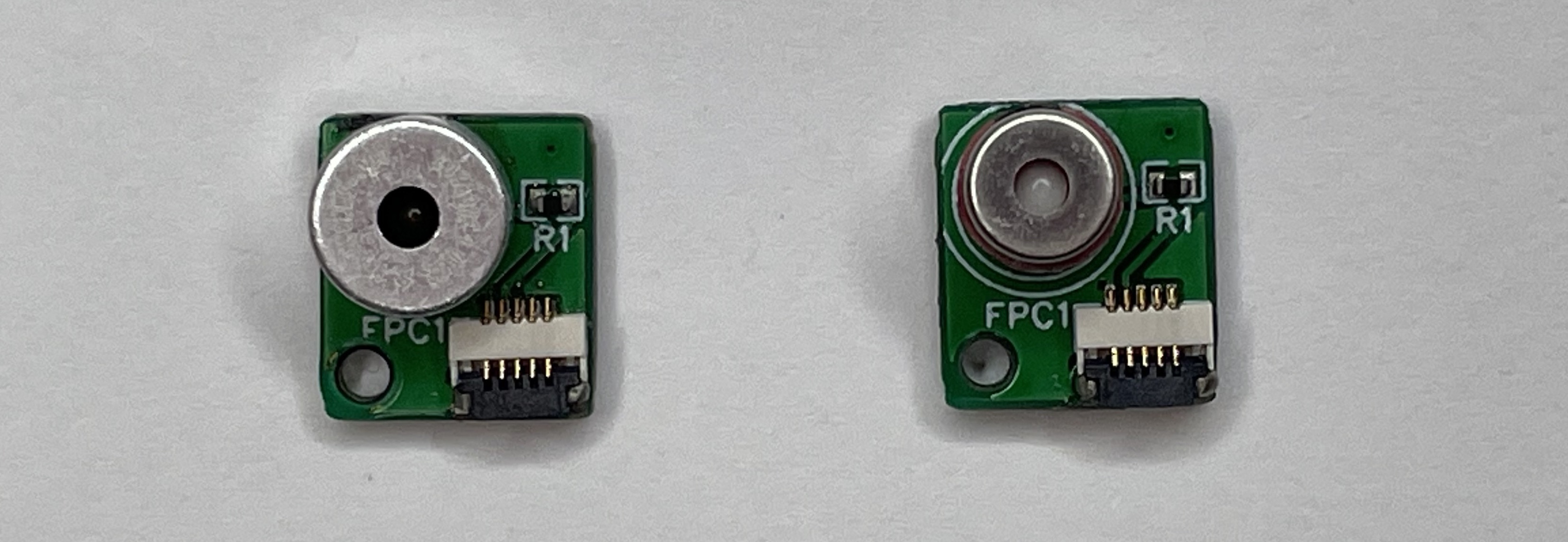}
  \caption{Comparison of Two Speakers: SR6438NWS-000 (Left) and OWR-05049T-38D (Right).}
  \label{Fig: different speakers}
  \Description{This figure shows two speakers side by side. They are soldered on the same type of PCB and very similar to each other. OWR-05049T-38D is slightly smaller than SR6438NWS-000.}
\end{figure}

As stated above, our power consumption is already relatively low, especially compared to using cameras for facial expression tracking. However, what we present in this paper is just a starting point. The power consumption of our system can be further optimized in the future. For instance, our measurement indicates that the two speakers take up about 80\% ($135mW$) of the system's power consumption ($167mW$). Therefore, depending on applications, EyeEcho, especially the speakers, do not need to be turned on all the time. Besides, reducing the loudness of the speakers and/or using high-efficiency speakers can also lead to lower power consumption. For instance, we replaced the two speakers in our system (SR6438NWS-000) with two speakers that are more power-efficient, OWR-05049T-38D \cite{owr} as shown in Fig.~\ref{Fig: different speakers}, and adjusted the Sound Pressure Level (SPL) to the same value as the previous speakers. Then we measured the current flowing through the system again and got the value $17.3 mA$ with the new speakers, which gave us a power consumption of $71 mW$. This validated that EyeEcho's power can be further reduced by adopting more power-efficient speakers.

\subsection{Real-time Deployment on Smartphones}
\label{Subsec: deployment on smartphone}
To demonstrate how EyeEcho can be integrated with commercial devices, we deployed the data processing and deep learning pipeline of EyeEcho with the help of PyTorch Mobile~\cite{pytorchmobile} on an Android smartphone (Xiaomi Redmi K40, Android 12, Qualcomm Snapdragon 870 SoC). Keeping the algorithms deployed on the smartphone will eliminate the need to transmit the data to a server on the cloud, which can better preserve the privacy of the user. 

\begin{figure}[h]
  \includegraphics[width=0.4\textwidth]{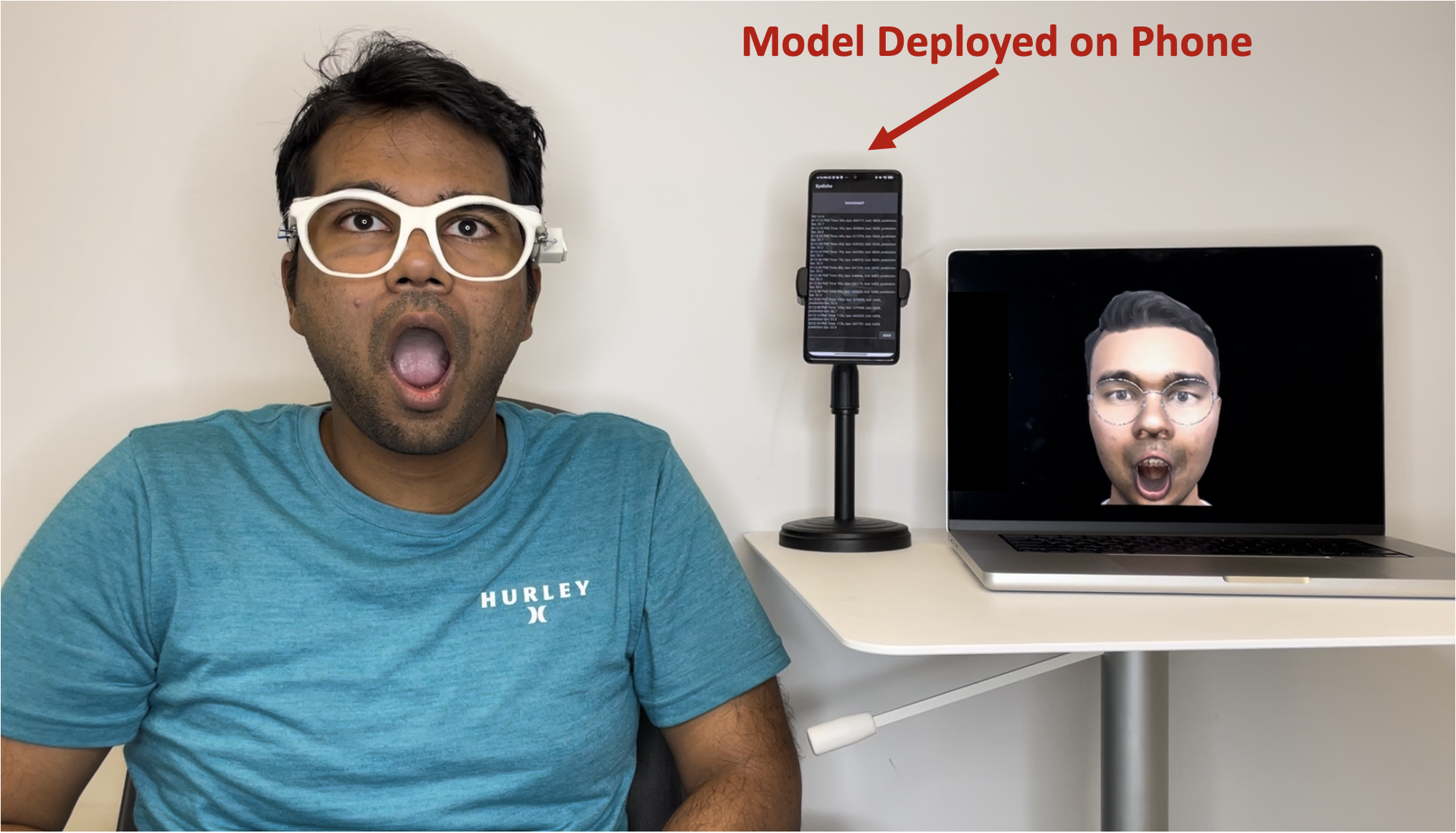}
  \caption{Real-time Pipeline Implemented on a Phone with Personalized Avatar Powered by Avaturn.}
  \label{Fig: real-time pipeline}
  \Description{This figure shows the real-time pipeline implemented on an Android phone. In the figure, a person is wearing the glasses prototype and doing the Open Mouth facial expression. The model deployed on the phone is making predictions in real-time. On the screen of a laptop, the personalized avatar is displaying the same facial expression that the person is performing.}
\end{figure}

In order to keep the algorithm making predictions fast enough in real-time, we replaced the deep learning model with a ResNet-18 architecture. To demonstrate that this lighter model results in a similar performance, we trained the lighter model (ResNet-18) in the same way as we did in Sec.~\ref{Subsec: user-dependent model} for the sitting scenario, achieving comparable performance with the full model (23.1 vs 22.9). For the usage of the real-time pipeline, we first trained the deep learning model with the data collected from users in PyTorch on a NVIDIA GeForce RTX 2080 Ti GPU. Then we traced the trained model to make it applicable to deployment on smartphones. The traced model was loaded onto the Android phone (Xiaomi Redmi K40) and used in the Android application we developed for data collection and facial expression prediction. During the inference stage, the running App continuously received data streamed from the BLE module in our EyeEcho system via Bluetooth, preprocessed the received data (i.e. organizing data based on channels, filtering the raw data, and calculating the echo frames), and fed echo profiles into the traced deep learning model for predictions. The predicted blendshape parameters representing users' facial expressions were finally streamed from the smartphone to a laptop via Wi-Fi for downstream applications. On average, it took $34 ms$ to make one inference, which led to a refresh rate of 29 FPS. We believe that this is sufficient for most applications since most videos can be played at 30 FPS. With the system implemented on the smartphone, we were able to predict users' facial expressions in real-time and rendered them with a personalized avatar powered by Avaturn\footnote{\url{https://avaturn.me/}}, as shown in Fig.~\ref{Fig: real-time pipeline}.

\subsection{Evaluating Blink Detection}
\label{Subsec: blinking detection}
Blinking is an important part of facial movements and can be used to monitor health conditions and help diagnose many eye diseases of users \cite{park2019dryeye,dementyev2017dualblink}. In the past, in order to detect blinks, separate sensors have been installed on glasses, such as cameras \cite{xiong2017iblink} and capacitive sensor \cite{liu2022noncontact}. To be best of our knowledge, there is no acoustic-based system that can detect blinks based on skin deformations around the cheeks.  

In our preliminary experiments, we noticed that blinks also lead to substantial deformations on tissue and skin around the face. This was visually evident in the echo profiles extracted from the received acoustic data. Consequently, we believe that, in addition to tracking facial expressions, our system has the capacity to detect blinks.

To evaluate the feasibility of EyeEcho for blink detection, we conducted a follow-up study with 6 participants. In this follow-up study, we asked each participant to watch some video clips of landscapes while wearing the EyeEcho system. We did not instruct them to blink intentionally. Instead, EyeEcho system detected the natural blinks of participants while watching these videos. While wearing the EyeEcho device, they watched five 2-min video clips. Between two videos, they remounted the device. For the five sessions of data, we ran a 5-fold cross-validation using 4 sessions to train the model and testing on the remaining session. During training, we only used 2 out of 52 blendshape parameters that are related to blinking (eyeBlink\_L/R) as the ground truth for the model to optimize blinking detection performance. The average F1 score of blinking detection across 6 participants was 82\% ($std=12\%$) for our system. It's important to note that EyeEcho was not specifically designed for blink detection, so performance could be further enhanced by incorporating additional sensors positioned closer to the eyes on glasses and oriented directly towards the eyes. This study aims to showcase the potential feasibility of using EyeEcho to simultaneously track facial expressions and blinks.

\subsection{Long-term Evaluation of the System}
\label{Subsec: long-term evaluation}
Over time, the prediction of the system could become worse because the user's body status is changing every day and will not be exactly the same as the day when the training data is collected. In the semi-in-the-wild study in Sec.~\ref{Subsec: wild study}, we validated that our system still works well on Day Two when training data is collected on Day One. In this subsection, we would like to conduct a more thorough long-term evaluation of the system. Thus two researchers collected the same amount of training data as we did in the in-lab user study and tested the performance of the system on the same day as well as 1 day, 2 days, 1 week and 2 weeks after the training data was collected. The average testing results are shown in Tab.~\ref{Tab: long-term results}. The average MAE across the two researchers are 19.9, 33.6, 40.4, 36.3 and 36.0 for the test data collected on the same day and 1 day, 2 days, 1 week and 2 weeks after the training data was collected. The result shows that the performance of our system decreases over time but is still good enough even after 2 weeks, based on analysis in Sec.~\ref{Subsec: evaluation metrics} and Sec.~\ref{Subsubsec: visualized results}, which proves the stability of our EyeEcho system.

\begin{table}[h]
\caption{Long-term Evaluation Results.}
\label{Tab: long-term results}
\begin{tabular}{| c | c | c | c | c | c |} 
\hline
 & 0 Day & 1 Day & 2 Days & 1 Week & 2 Weeks\\ [0.5ex] 
 \hline\hline
 MAE & 19.9 & 33.6 & 40.4 & 36.3 & 36.0\\
 \hline
\end{tabular}
\end{table}

\begin{table*}[h]
\caption{Comparison with EarIO in MAE and Training Data Needed.}
\label{Tab: comparison eario}
\begin{tabular}{| c | c | c | c | c | c |} 
\hline
 Project & Training Data Needed & Performance in Sitting Scenario & Performance in Walking Scenario\\ [0.5ex] 
 \hline\hline
 \multirow{2}{*}{EyeEcho} & 4 Minutes & 29.7 (MAE) & 34.3 (MAE)\\
 \cline{2-4}
  & 20 Minutes & 22.9 (MAE) & 26.9 (MAE)\\
 \hline
 EarIO~\cite{li2022eario} & 32 Minutes & 25.9 (MAE) & 33.9 (MAE)\\
 \hline
\end{tabular}
\end{table*}

\subsection{Comparison with EarIO}
\label{Subsec: comparision with eario}

\subsubsection{Different form factors and sensor positions to track different movements} 
EarIO~\cite{li2022eario} bears close relevance to EyeEcho as both utilize active acoustic sensing for continuous facial expression tracking. However, EarIO places sensors on a pair of earphones with a 3D-printed attachment to monitor movements beneath the ears, whereas EyeEcho enables facial expression tracking on glasses with sensors placed on the legs of the glasses to capture skin deformations around the eyes and cheeks. These differing hardware configurations result in distinct performance and limitations. Furthermore, the use of earphones differs significantly from that of glasses. While many people are used to wearing glasses throughout the day, most people may not be comfortable wearing earphones for daily activities. Therefore, even though earphones have the capability to track facial expressions, it remains essential to explore the tracking of facial expressions on glasses.

To enable a comprehensive and rigorous comparison between EyeEcho and EarIO, we reproduced the EarIO system, including algorithms and data and conducted a side-by-side evaluation.

\subsubsection{Better performance with less training data} 
First, the full user study of EyeEcho and EarIO employed the same ground truth acquisition method (TrueDepth camera) and evaluation metrics (MAE). Based on the results displayed in Tab.~\ref{Tab: comparison eario}, EyeEcho with 20 minutes of training data outperforms EarIO with 32 minutes of training data in both static settings (22.9 vs 25.9 in MAE) and mobile settings (26.9 vs 33.9 in MAE), indicating improved robustness. In the meantime, if EyeEcho only uses 4 minutes of training data, it still achieves comparable performance to EarIO which uses 32 minutes of training data (29.7 vs 25.9 for sitting and 34.3 vs 33.9 for walking). This demonstrates that to obtain a similar tracking performance, EyeEcho only requires 12.5\% training data compared with EarIO (4 mins vs 32 mins).

\subsubsection{New ability to detect blinking} As described in Sec.~\ref{Subsec: blinking detection}, EyeEcho achieves an F1-score of 82\% for blinking detection across different sessions. EarIO did not conduct studies on blinking detection. Therefore, we replicated the sensing system of EarIO and implemented similar blinking detection algorithms from our system. To compare two systems side by side, three researchers and one participant evaluated blinking detection using both EyeEcho and EarIO with a similar study procedure and setup as Sec.~\ref{Subsec: blinking detection}. To better demonstrate the best performance of each system for blinking detection, we collected data within one session for this experiment. Results show that EyeEcho can detect blinks with an F1 score of 99\% while EarIO achieved an F1-score of 0\% even within one session across these four people. We believe this is because the movements behind the ear and chin, which EarIO captures, are not sensitive to subtle eye movements compared to the skin deformations around the cheeks, captured by EyeEcho.

\subsubsection{Better stability over a long period of time} In addition, we also tested the facial expression tracking performance of both systems in a longitude study. EyeEcho maintains consistent performance even after 2 weeks, as illustrated in Sec.~\ref{Subsec: long-term evaluation}. By contrast, tested by one researcher, the performance of EarIO degrades significantly from 20.5 to 47.1, 42.7, and 51.2 in MAE when tested 2 hours, 12 hours and 1 day after training data was collected. It indicated that our system can have better stability if being deployed in a long period of time. We think this is because the glass frame is more stable compared to earables. For instance, the wearing position of glass frames is relatively consistent while the wearing position of earables can shift after each remounting session.

In summary, the results from these preliminary studies showed that EyeEcho outperforms EarIO~\cite{li2022eario} in terms of performance, the training data needed, the ability to detect blinks, and stability. Moreover, EyeEcho includes a semi-in-the-wild evaluation, demonstrating consistent performance in naturalistic settings, while EarIO was only evaluated in controlled lab settings.

\subsection{Health Implications}

In Sec.~\ref{Subsec: frequency impact}, we validated that EyeEcho can operate in the frequency range $20-24 kHz$, which is inaudible to users. However, even though the users cannot hear the signal, it may still cause health concerns to them. Thus, we used the NIOSH Sound Level Meter App\footnote{\url{https://www.cdc.gov/niosh/topics/noise/app.html}} to measure the signal level of the EyeEcho system. We kept the speakers in the system emitting the signals and attached the microphones of the phone with the NIOSH app running directly onto the speakers. The average sound level measured was 48.2 dB. When we moved the microphones of the phone to the same distance from the speakers of EyeEcho as where the users' ears will be if they wear the system, the measurement of the sound level was 37.8 dB. According to Howard et al.~\cite{howard2005review}, the recommended ultrasound exposure limit for frequency around $20 kHz$ is 75 dB. Therefore, we believe that our EyeEcho system is safe to wear for long-term use since the sound level is far from the recommended limit.

\subsection{Privacy Preservation Mechanisms in EyeEcho} 

EyeEcho can preserve privacy by avoiding the use of cameras and limiting the frequency range of the sensing system. 
EyeEcho uses a band-pass filter to remove all frequencies other than $16-20 kHz$ in the received signal, which means that all audible sounds including heavy privacy information (e.g., human speaking, environmental sounds) are removed.
Furthermore, we also demonstrated that EyeEcho can conduct all computations locally on a smartphone without sending any data over the Internet in Sec.~\ref{Subsec: deployment on smartphone}. In this way, sensitive information stays confidential because only the predicted facial expressions represented by 52 parameters instead of the original signals captured by microphones are shared with others. With future advancement in low-power on-chip deep learning, it is also possible to deploy everything on a single chip, further eliminating the privacy and security risks.

\subsection{Applications on Commodity Devices}
\label{Subsec: applications}

Enabling facial expression tracking on glasses have a wide range of applications, from enhancing video conferencing experience to novel input methods. Video conferencing, a common mode of communication, can be significantly improved using our system. Currently, during online video conferences, participants must position themselves in front of a camera or hold it to ensure others can see their facial expressions. However, with EyeEcho integrated into smart glasses, users can engage in real-time video conferences and convey facial expressions effortlessly, even while walking or multitasking. We have implemented a system capable of generating personalized avatars with facial expressions for each user, as detailed in Sec.~\ref{Subsec: deployment on smartphone}. Consequently, the user experience in video conferencing with EyeEcho resembles traditional camera-based methods, but with the added convenience of hands-free operation.

Facial expressions can also serve as a novel input method, a concept has been explored in previous research efforts~\cite{rantanen2010capacitive,lankes2008facial,matthies2017earfieldsensing,Mei2018a}. Our glasses-based system is well-suited for implementing this functionality, allowing different facial expressions to serve as distinct commands for interacting with smart or augmented reality glasses.

Moreover, facial expressions are linked to various health conditions. For instance, individuals with Parkinson's disease may experience a loss of facial expressions. EyeEcho can be instrumental in tracking and monitoring the symptoms of such diseases, potentially contributing to improved healthcare outcomes.

\subsection{Limitations and Future Work}

Despite the promising performance, EyeEcho also has limitations that need further investigation.

\subsubsection{Impact of Vigorous Exercises}
The system performance might be negatively impacted if the user conducts vigorous exercises (e.g., shaking heads, and running). This can potentially be alleviated if we improve the form factor design and collect training data from these sessions.

\subsubsection{More Diverse Evaluation Settings and Environments}
We only evaluated the system when the participants were performing a selected set of facial expressions in our in-lab study. In the semi-in-the-wild study, we only conducted the study in an apartment. The goal of this paper is to demonstrate the feasibility of the first acoustic continuous facial expression tracking system on glasses. We plan to assess the system in more daily settings and environments (e.g., offices, classrooms, dining halls) in the future. 

\subsubsection{Environmental Impact}
Although we validated that our system works equally well at the frequency range of $20-24 kHz$, which is inaudible to most people, it may still be heard by certain people (especially kids) and animals. We will conduct experiments to understand how our system impacts the environment and choose frequency and sensors accordingly.

\subsubsection{Impact of Real-world Factors}
Even if it did not happen in our user study, it is possible that certain types of long hair can cover the sensors which may lead to the failure of the sensing system. People with heavy beards on their cheeks may also encounter problems while using this system. This can be a limitation of this sensing system.

\subsubsection{Explore More Sensor Positions}
While prototyping EyeEcho, we explored three sensor positions on glasses which we thought are most likely to deploy sensors and picked the one that achieved the best performance and obtrusiveness. We plan to explore more possible positions on glasses to determine the optimal one for our system.
\section{Conclusion}
\label{Sec: Conclusion}

This paper introduces EyeEcho, a low-power and minimally obtrusive technology designed for glasses that enables continuous facial expression tracking. It represents the first successful implementation of on-device acoustic sensing for tracking facial expressions continuously. The system's capabilities were assessed through both in-lab and semi-in-the-wild studies, revealing promising performance across diverse scenarios. Additionally, we successfully demonstrated the system can be deployed on an off-the-shelf smartphone for real-time processing.  These outcomes underscore the significant potential for EyeEcho to be integrated into future smart glasses for real-world applications.

\begin{acks}
This work is supported by National Science Foundation (NSF) under Grant No. 2239569, NSF's Innovation Corps (I-Corps) under Grant No. 2346817, NSF Award IIS-1925100, the Ignite Program at Cornell University, the Nakajima Foundation, and the Ann S. Bowers College of Computing and Information Science at Cornell University. We also appreciate the help of our lab mates on providing feedback on the hardware prototypes and paper writing. Specifically, we would like to thank Devansh Agarwal and Jian Wang for their efforts on developing the real-time demonstration of the system.
\end{acks}

\bibliographystyle{ACM-Reference-Format}
\bibliography{sample-base}

% \input{11_appendix}


\end{document}